\documentclass[fleqn,usenatbib]{mnras}

\usepackage{newtxtext,newtxmath}

\usepackage[T1]{fontenc}
\usepackage{ae,aecompl}

\usepackage{graphicx}	
\usepackage{amsmath}	
\usepackage{amssymb}	
\usepackage{ tipa }
\hypersetup{draft}
\usepackage{xtab,afterpage}
\usepackage{longtable}

\def\teff{$T_{\rm eff}$}

\title[Characterizing  \textit{TESS} host stars]{Know thy star, know thy planet: Chemo-kinematically characterizing TESS targets}

\author[A. Carrillo et al.]{
Andreia Carrillo$^{1,2}$\thanks{E-mail: andreiac@utexas.edu},
Keith Hawkins$^{1}$,
Brendan P. Bowler$^{1}$,
William Cochran$^{1}$,
\newauthor
and Andrew Vanderburg$^{1}$
\\
$^{1}$Department of Astronomy, University of Texas at Austin, 2515 Speedway, Stop C1400, Austin, TX 78712-1205, USA\\
$^{2}$Large Synoptic Survey Telescope Corporation Data Science Fellow
}

\date{Accepted XXX. Received YYY; in original form ZZZ}

\pubyear{2019}

\begin{document}
\label{firstpage}
\pagerange{\pageref{firstpage}--\pageref{lastpage}}
\maketitle

\begin{abstract}
The Transiting Exoplanet Survey Satellite (\textit{TESS}) has already begun to  discover what will ultimately be thousands of exoplanets around nearby cool bright stars. These potential host stars must be well-understood to accurately characterize exoplanets at the individual and population levels. We present a catalogue of the chemo-kinematic properties of 2,218,434 stars in the \textit{TESS} Candidate Target List using survey data from $Gaia$~DR2, APOGEE, GALAH, RAVE, LAMOST, and photometrically-derived stellar properties from SkyMapper. We compute kinematic thin disc, thick disc, and halo membership probabilities for these stars and find that though the majority of \textit{TESS} targets are in the thin disc, 4\% of them reside in the thick disc and $<$1\% of them are in the halo. The \textit{TESS} Objects of Interest in our sample also display similar contributions from the thin disc, thick disc, and halo with a majority of them being in the thin disc. We also explore metallicity and [$\alpha$/Fe] distributions for each Galactic component and show that each cross-matched survey exhibits metallicity and [$\alpha$/Fe] distribution functions that peak from higher to lower metallicity  and lower to higher [$\alpha$/Fe] from the thin disc to the halo. This catalogue will be useful to explore planet occurrence rates, among other things, with respect to kinematics, component-membership, metallicity, or [$\alpha$/Fe].
\end{abstract}




\section{Introduction}
The search for exoplanets necessitates the detailed characterization 
of planet-hosting stars to map the diversity of exoplanets and ultimately understand the Galactic context of planet formation. 
Stars and their planets are formed from the same material, so determining the ages, compositions, and kinematics of planet-hosting stars is necessary to establish how, when, and under what conditions rocky planets and gas giants are created. 

Within the last decade, the study of exoplanets has been very fruitful through various methods.
The transit method 
has proven to be a particularly effective tool to discover exoplanets with dedicated space telescopes such as the  \textit{Kepler} spacecraft \citep{borucki10}, which was designed to search for planets surrounding Sun-like stars, as well as the recently launched \textit{Transiting Exoplanet Survey Satellite} (\textit{TESS}, \citealt{ricker15}). Follow-up high spatial imaging and precision radial velocities 
(e.g., \citealt{batalha11,borucki12,barclay13,borucki13}), enable further validation and characterization of the planetary systems. 
The same data also provide information about host stars' physical properties i.e. effective temperature (\teff), surface gravity (log~$g$), spectral type, metallicity ([Fe/H]
, and chemical abundances ([X/Fe]).

Exoplanet discoveries over the past quarter century have opened doors to studying the diversity of planet populations and how they relate to the properties of their host star \citep{mulders18}. There is an observed  higher occurrence of short-period gas giants around more metal-rich host stars \citep{gonzalez97,fischervalenti05,johnson10}, 
providing evidence of the core-accretion model of planet formation where the cores of giant planets form more rapidly when more metals and more massive planetary discs are available \citep{pollack96,petigura18}. This is not observed for Earth and Neptune-mass planets \citep{sousa08} that seem to occur around stars with a large range of metallicities \citep{buchhave12, petigura18}. 
Planets smaller than three times Earth's radius occur more frequently around M dwarfs than FGK stars, but this relationship inverts for gas giants which are twice as common around FGK stars than M dwarfs \citep{howard12, mulders15c}. Decreasing host star $ T_{\rm eff}$ also correlates with higher planet occurrence \citep{howard12}.
With these studies, it has become clearer that in order to find more Earth-like planets, we have to search for them around cooler and smaller stars.

These statistics have helped motivate the new generation planet finder satellite and successor to  \textit{Kepler}, \textit{TESS}. 
\textit{TESS} is actively obtaining light curves for $\sim$ 200,000 main-sequence dwarf stars at 2-minute cadence as well as full frame images (FFI) for most of the sky every 30 minutes. It is targeting stars 30-100 times brighter than \textit{Kepler} host stars to allow for easier follow-up and will cover 85\% of the sky by the end of its primary mission. Simulations by \citet{barclay18} predict that \textit{TESS} will find over 14,000 exoplanets, 2100 of which would be smaller than Neptune and 280 of those would have radii less than 2$R_{\oplus}$. The \textit{TESS} 2-minute cadence and FFI observations will include a broad range of host stars in the different Galactic components, the thin disc, thick disc, and halo. This will help improve the link between the types of planets that exist around host stars with different properties and environments. 

Characterization of the Galactic components has been carried out in both kinematic space and chemical space, for a multitude of stars in the solar neighborhood and to larger galactocentric radii, to study the thin disc, thick disc, and halo of the Milky Way \citep{freeman02,rix13,hayden15}. The thin disc is vertically thin \citep{bovy12b,haywood13}, rotation dominated \citep{edvardsson93,rix13}, has a metallicity range of $\rm -0.7 < [Fe/H] < +0.5$ dex \citep{bensby14}, and has ongoing star formation. The thick disc \citep{yoshii82,gilmore83} has larger vertical scale-heights but smaller radial scale-lengths than the thin disc \citep{bovy12b,haywood13,hayden15}, and has rotation but also higher velocity dispersion \citep{haywood13,kordopatis13}. It is also chemically different than the thin disc, having lower metallicities and higher [$\alpha$/Fe] \citep{edvardsson93,bensby03,kordopatis13, nidever14}, pointing to a scenario where it is older than the thin disc. Alpha elements (O, Mg, Si, S, Ca) are dispersed into the interstellar medium due to core-collapse supernova (SN II). However, [$\alpha$/Fe] goes down with time as supernova Type Ia (SNe Ia) events start to happen that contribute more Fe. Stars with enhanced [$\alpha$/Fe] are formed from gas that was enriched primarily by core-collapse supernovae. Therefore, [$\alpha$/Fe] is indicative of a stellar populations' star formation history, with stars having high [$\alpha$/Fe] being older, on average. 

Lastly, the halo is comprised of stars accreted from satellites and stars formed from rapid gas collapse during the Galaxy's infancy \citep{eggen62,searle78,ibata94,belokurov06}. These stars are found to be very old and metal-poor \citep{mcwilliam97,jofre11,hawkins14}. It could be further subdivided into an inner halo and outer halo that separate in space, kinematics, and metallicity \citep{carollo10}. Compared to the thin and thick discs, the halo has lower metallicities, higher [$\alpha$/Fe], and is more pressure-supported \citep{mcwilliams95,carollo10,ishigaki12}.

Multiple astrometric, spectroscopic, and photometric surveys have improved our understanding of the Galaxy and its different components. We have positional and photometric information for 1.38 billion stars as well as radial velocities for 7 million stars brought by the $Gaia$ mission \citep{gaiasatellite}, allowing us to make the most precise 3D map of the Milky Way to date. Large spectroscopic surveys have also been utilized to study the chemical make-up and cartography of the Milky Way. These surveys include the high-resoultion infrared Apache Point Observatory (APO) Galactic Evolution Experiment (APOGEE; \citealt{majewski17}), which samples hundreds of thousands of evolved stars in the bulge, disc, and halo, and the optical Galactic Archaeology with Hermes (GALAH; \citealt{desilva15}) survey, which provides chemical abundances for 23 elements of very local dwarfs and giant stars. Other large spectroscopic surveys at lower resolution provide information for even more stars. These include the optical Radial Velocity Experiment (RAVE; \citealt{kunder17}), which aims to measure precise radial velocities for hundreds of thousands of dwarf and giant stars in the southern hemisphere and the Large Sky Area Multi-Object Fiber Spectroscopic Telescope (LAMOST; \citealt{cui12}), which provides stellar parameters for 5 million stars in the northern hemisphere. Each survey is important because they bring forth stellar paramaters and metallicity (for some, also chemical abundances) for a statistically large sample of stars in various parts of the Galaxy, derived from the optical to the infrared. 

Providing stellar parameters for even more stars are photometric surveys such as SkyMapper \citep{wolf18}, which has observations for 285 million stars in the Southern hemisphere. This has been utilized by \citet{casagrande19} and \citet{deacon19} to derive stellar parameters such as \teff~and [Fe/H]. These surveys enable studies of the properties (\teff~and [Fe/H]) of orders of magnitude more stars across the Galaxy than spectroscopic surveys. 

We are now capable of studying the chemical and kinematic structure of the Milky Way in great detail, and with \textit{TESS} discovering thousands of exoplanets in the coming years, it is imperative that we study these populations in their Galactic context and understand their host stars which is the goal of this study. Therefore, this paper is organized as follows: In Section~\ref{sec:data} we outline the details of the surveys we have used to perform the chemical and kinematic analysis, in Section~\ref{sec:results} we discuss the kinematic and chemical properties of 2,218,434 \textit{TESS} host stars using the various photometric, astrometric, and spectroscopic surveys. We then tie together the chemistry and kinematics of \textit{TESS} targets from the Candidate Target List (CTL) and TESS Objects of Interest (TOI) in Section~\ref{sec:discussion}. The results of this work are summarized in Section~\ref{sec:summary}. We also provide the column names for the catalogue produced from this study (Table \ref{tab:catalogue}).

\section{Data}
\label{sec:data}


\subsection{TESS}
\label{sec:tess}

We use the  \textit{TESS} CTL v8.01~\footnote{\url{https://filtergraph.com/4701718}}, a refined list of targets made from the  \textit{TESS} Input Catalog (TIC) version 8  \citep{stassun19}. This version of TIC uses \textit{Gaia} DR2 as base for better positions and parallaxes and as a result contains a total of $\sim$1.7 billion point sources (See Appendix A1 for a comparison and discussion of TICv7). The  expected \textit{TESS}  magnitude was calculated using relations with the \textit{Gaia} $G$, $G_{BP}$, and $G_{RP}$ photometry. The spectroscopically-determined \teff~values (from catalogues listed in their Table 1) were provided if available, but otherwise, the photometrically-determined \teff values were given. The \teff~ is used to calculate radii, mass, and log $g$ for the sources in TIC. The metallicity for an object is provided in the TIC if the star is cross-matched with the spectroscopic catalogues in their Table 1; for cases where there is more than one metallicity, the values are combined using a weighted-mean. Although the metallicity is included in the TIC, we independently cross-match the TESS targets with the individual spectroscopic and photometric surveys in order to avoid combining data (e.g., metallicity) with different selection effects.  

The CTL was generated from a subset of the TIC as well as several curated lists including cool dwarfs (\citealt{miurhead18}, Muirhead et al., in prep.), hot subdwarfs \citep{geier17}, and Guest Investigator Cycle 2 targets. The catalogue serves as a list of potential \textit{TESS} targets for the 2-minute cadence to (1) search for planets with periods of less than 10 days and radii less than 2.5$R_{\oplus}$, (2) search for planets with radii less than 2.5$R_{\oplus}$ and longer periods of 120 days for targets near the ecliptic poles and (3) to deliver masses for 50 planets with radii less than 4 $R_{\oplus}$.  After the distillation of TIC to only include cool bright dwarfs and remove evolved stars using the radius derived for TIC sources, the CTL contains a total of 9.48 million targets.   The stars in the CTL also fulfill the following cuts: (1) parallaxes and \textit{Gaia} photometry satisfy equations 1 and 2 from \citealt{arenou18} (cuts made to only include sources with good astrometric solutions)  and (2) $T$ < 13 mag in order to reduce the CTL to a manageable size while prioritizing bright dwarfs.

The CTL provides a tabulated priority value (that goes from zero to one with one being the highest) which is used to optimize the detection of a light curve with \textit{TESS}. The priority value depends on the stellar radius, expected photometric precision, and number of sectors a source will be observed in. For example, a star that falls on multiple \textit{TESS} sectors will have boosted priority.  High priority is also given to stars in the special curated lists. On the other hand, stars are de-prioritized if they are near the Galactic plane with $|b| < 10^{\circ}$ because of poorly-understood reddening effects. 

\subsection{Gaia DR2}
\label{sec:gaiadr2}

\begin{figure}
\center
\includegraphics[width=0.4\textwidth]{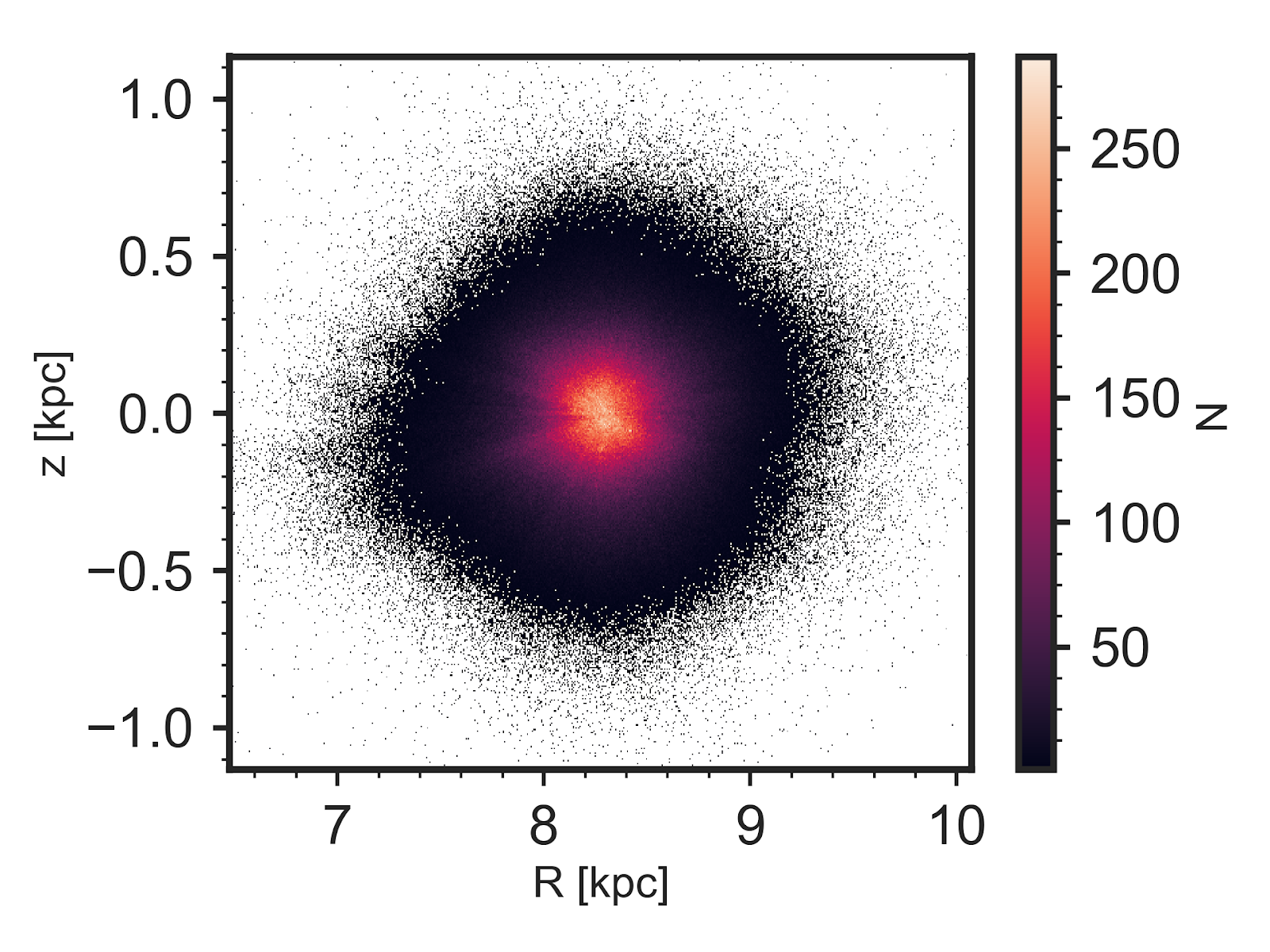}
\caption{ Galactocentric radius (R) vs height (z) for the TG sample using distances from \citet{bailerjones18}, coloured by the density. (0,0) marks the Galactic center. }
\label{fig:Rz}
\end{figure}

\begin{figure}
\center
\includegraphics[width=0.4\textwidth]{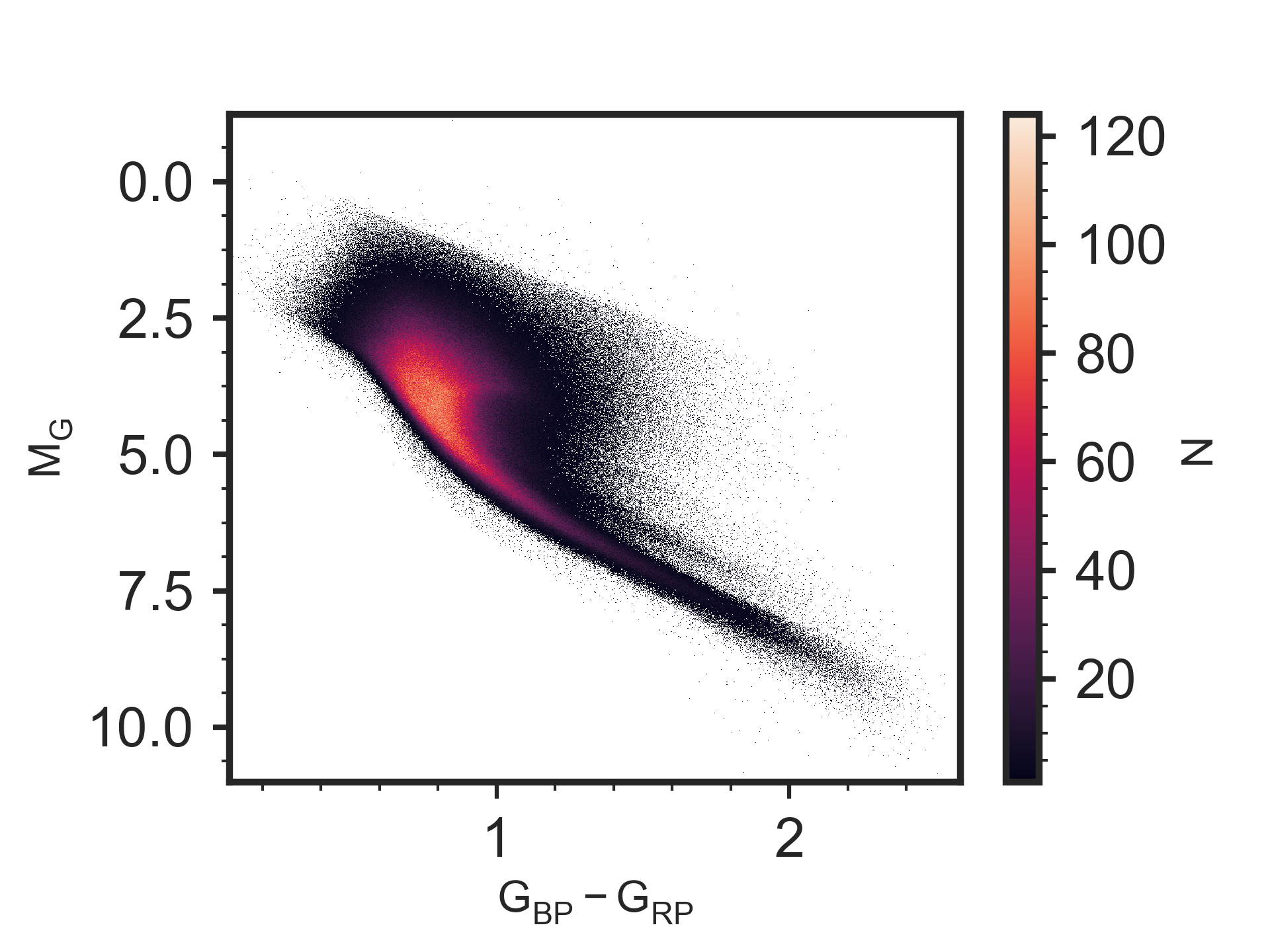}
\caption{Color-magnitude diagram of the  TG sample that shows that most of the TESS targets are main sequence stars.} 
\label{fig:cmd}
\end{figure}

The European Space Agency satellite $Gaia$ \citep{gaiasatellite} aims to paint a precise three-dimensional picture of our Galaxy to establish its structure and evolution.  $Gaia$'s most recent data release (DR2, \citealt{gaiadr2}) includes 
radial velocities for 7.2 million stars which was used in \citet{marchetti18} to derive full 6D phase-space information in search of unbound stars in the Galaxy. 

We use their kinematic catalogue for this study, specifically their derived UVW velocities. Since CTL v8 uses \textit{Gaia} DR2 as base, it was straightforward to cross-match with the \citet{marchetti18} catalogue. We made quality cuts in the $Gaia$ BP and RP photometry by removing the sources without these values, and applied the cut \texttt{Gaia\_astrometric\_excess\_noise\_sig $ \leq 2$} following \citet{marchetti18}, to ensure that we only select sources that are astrometrically well-behaved. We note that this cut preferentially deselects close or unresolved binaries \citep{evans18}.  We also applied a relative error cut to the total velocity,\textit{ v}, such that $\sigma_{\rm v}/\rm v < 0.3$. This final sample, which we now refer to as ``TG" (for \textit{TESS} cross-matched with $Gaia$), contains 2,218,434 sources.The decrease in the number of sources from 9.5 million to 2.2 million is due to the requirement of a \textit{Gaia} radial velocity measurement. The net effect of this selection biases the TG sample to only include stars brighter than  \textit{G} $\sim$14 mag \citep{gaiadr2}.

The location of stars in the TG sample in the Milky Way is shown in Figure \ref{fig:Rz}, where the Galactocentric radii and heights are derived using the distances from \citet{bailerjones18}, which are computed in a Bayesian framework using a weak distance prior that varies as a function of Galactic position from $Gaia$ DR2. Other distance catalogues also exist (e.g., \citealt{queiroz18,schoenrich19,leung19}) but we choose \citet{bailerjones18} to be consistent with the analysis of \citet{marchetti18}. Figure \ref{fig:cmd} shows a 2D histogram of the TG colour magnitude diagram (CMD) showing that most of the targets are main sequence stars. 

\subsection{APOGEE}
\label{sec:apogee}

In addition to the astrometric and inferred kinematic data, we also take advantage of spectroscopic data from The Apache Point Observatory (APO) Galactic Evolution Experiment (APOGEE; \citealt{holtzman18}) DR14 that has 258,475 red giants stars and evolved stars. APOGEE has moderate spectral resolution (R $\sim$ 22,500) data taken in the \textit{H}-band (1.5-1.7 $\mu$m), using the Sloan Foundation 2.5-m Telescope at APO. It uses fiber-plugged plates with a maximum simultaneous observation of 300 fibers for an area of 1.0 $\rm deg^2$, enabling the acquisition of many stellar spectra at the same time. Since APOGEE is in the near-infrared, it is less sensitive to the effects of dust. 

The spectra have been used to derive stellar parameters (log $g$, \teff~, microturbulence, [Fe/H]) and abundances through the  APOGEE Stellar Parameters and Chemical Abundances Pipeline (ASPCAP; \citealt{garciaperez16}). ASPCAP determines the best-matching synthetic spectra \citep{zamora15} with known stellar parameters to the observed spectra. 
We cross-match APOGEE with the TG sample using the 2MASS ID and applying the following quality control cuts: (1) STARFLAG = 0 to ensure no warnings on the observation, (2) ASPCAPFLAG = 0 to only select stars whose parameters have converged and have no warning flags, (3) [Fe/H] error \textdoublebarslash -9999, (4) signal-to-noise (SNR) $>$ 80 to ensure high SNR, and (5) 4000 $< T_{\rm eff} <$ 5500 K because estimates of ASPCAP outside this range are less reliable. This yields a final APOGEE-TG sample of 658 sources. We use the [Fe/H] and [$\alpha$/Fe] in this study.

\subsection{GALAH}
\label{sec:galah}
In addition to APOGEE, we also use the Galactic Archaeology with Hermes (GALAH; \citealt{martell17}; \citealt{buder18}) survey. GALAH derives stellar parameters and chemical abundances for 342,682 stars in the Milky Way. It is a rich, high resolution (R $\sim$ 28,000) data set that enables us to understand the evolution of the Galaxy through a comprehensive list of abundances that range from light to neutron capture elements. It has four discrete wavelength channels within the range 4710-7890~\AA\ taken with the HERMES Spectrograph at the 3.9-m Anglo-Australian Telescope. GALAH focuses primarily on stars in the disc, where most of the Milky Way's stellar mass resides. Stellar parameters were determined in a two-stage process. First, $\sim$10,000 stars are used as training set, with stellar properties (e.g., log $g$, \teff, microturbulence, vsin$i$, [Fe/H], and chemical abundances) determined through \textit{Spectroscopy Made Easy} v360 (SME, \citealt{valenti96,piskunov17}). Then these derived properties are used as labels to determine the stellar parameters and abundances for the rest of the GALAH sample using \textit{The Cannon} \citep{ness15}. 

For this study, we utilize GALAH DR2 released in April 2018 and the \textit{Gaia} source ID to cross-match with the TG sample. We apply the following quality cuts: (1) flag\_cannon = 0, (2) flag\_x\_fe = 0 for $\alpha$ elements to ensure no warnings and make sure we have reliable derived parameters, (3) SNR $>$ 20, and (4) 4000 $< T_{\rm eff} <$ 7000 K as the GALAH pipeline is not optimized for stars outside this range. Our meta-catalogue incorporates the [Fe/H], [$\alpha$/Fe], log $g$, and $ T_{\rm eff}$ from the final GALAH-TG sample containing 32,517 sources.


\subsection{RAVE}
\label{sec:rave}
The fifth data release of the Radial Velocity Experiment (RAVE; \citealt{kunder17}) is a magnitude-limited (9 mag < \textit{I} < 12 mag) spectroscopic survey with the goal of measuring precise radial velocities of stars with an accuracy of 1.5 km$\rm s^{-1}$ as well as deriving \teff, log $g$, [Fe/H], and [X/Fe] (for X = Mg, Al, Si, Ti, Ni) for 457,588 randomly-selected stars in the southern hemisphere. It has a medium resolution of R$\sim$7,500 with a wavelength range spanning 8410-8795~\AA\, which includes the Calcium triplet. Observations were taken at the  1.2m UK-Schmidt Telescope by the Australian Astronomical Observatory. 

RAVE DR5, made available in November 2016, has improved surface gravities for giants and distances for metal-poor stars compared to RAVE DR4 \citep{kordopatis13rave} because of calibrations with asteroseismic data. The stellar parameters in RAVE DR5 were derived with the same pipeline as RAVE DR4: DEGAS (DEcision tree alGorithm for
AStrophysics, \citealt{bijaoui10}) for the low SNR spectra and MATISSE (MATrix Inversion for Spectral SynthEsis, \citealt{recioblanco06}) for the high SNR spectra. The RAVE data has also been re-analyzed with \textit{The Cannon} (RAVE-on, \citealt{casey16}) trained on the APOGEE data set which does not include many dwarfs. We therefore chose to use the stellar parameters derived from RAVE DR5 to encompass the main TESS targets. 

We use the 2MASS ID in the RAVE catalogue to cross-match with the TG sample. In doing so, the following quality cuts are applied: 1) Algo\_Conv\_K = 0 to ensure that the stellar parameter pipeline has converge, 2)  SNR $>$ 20, 3)  c1, c2, c3 (spectra morphological flags) are n as prescribed in \citet{kunder17}, 4) and Alpha\_c > -9.99. This yields  59,984 sources for the final cross-matched sample. We use [Fe/H],[$\alpha$/Fe], log $g$, and $T_{\rm eff}$ from this catalogue. 

\subsection{LAMOST}
\label{sec:lamost}
The Large Sky Area Multi-Object Fiber Spectroscopic Telescope (LAMOST; \citealt{cui12}) is a 4.0-m reflecting Schmidt telescope at the Xinglong Observatory, northeast of Beijing, China. LAMOST has a 5 deg$^{2}$ field-of-view and aims to observe spectra for 10 million stars, galaxies and quasi-stellar objects (QSOs) in the span of 5 years. At a resolution of R$\sim$1800 spanning 3900-9000 \AA, it is able to take spectra for 4000 objects in one exposure to a limiting magnitude of \textit{r}=19. LAMOST DR5 (released June 2019) has obtained spectra for 5,348,712 stars in the northern hemisphere and produced a catalogue of heliocentric radial velocities, \teff, log~$g$ and [Fe/H] for these stars. LAMOST is targeting three main regions: the Galactic anti-center, the disc, and the halo.

Stellar parameters are derived using the LAMOST Stellar Parameter Pipeline (LASP, \citealt{wu11,luo12}) which implements the Correlation Function Initial (CFI) method to initiate an original guess for the stellar properties that are then run through the University of Lyon Spectroscopic analysis Software code (UlySS, \citealt{koleva09}). The catalogue contains \teff, log $g$, [Fe/H], and radial velocity. There is also a Value Added Catalogue \citep{xiang19} derived with a data-driven approach that combines the capabilities of both \textit{The Payne} \citep{ting19} and \textit{The Cannon} \citep{ness15}. This catalogue includes 16 elemental abundances including [$\alpha$/Fe]. We cross-match LAMOST with the TG sample using a 5" radius between the LAMOST and $Gaia$ sky positions. We use the log $g$ and $T_{\rm eff}$ from the main LAMOST catalogue and the [Fe/H] and [$\alpha$/Fe] from the Value Added Catalogue after making the following quality cuts: removing sources with stellar parameters equal to -999, requiring that there be an [$\alpha$.Fe] value,  and imposing a SNR cut of $>$20 for \textit{griz}. The final cross-matched list has 344,565 sources. 

\subsection{Casagrande 2019}
\label{sec:casagrande}
\citet{casagrande19} derived \teff~and [Fe/H] for 9,033,662 stars in the Southern sky with the SkyMapper photometric survey \citep{wolf18}. SkyMapper has photometry for 285 million objects, taken on a 1.35m wide field survey telescope with 32 CCDs in Siding Spring Observatory, Australia. They use the $griz$ Sloan Digital Sky Survey filters \citep{fukugita96}, and SkyMapper filters $u$ and $v$ that are sensitive to hot stars and metallicity. \citet{casagrande19} determined the photometric zero point for SkyMapper using the InfraRed Flux Method (IRFM, \citealt{casagrande10}), which relates the ratio between bolometric and infrared flux to \teff. This method requires multiple photometric bands in the optical and the infrared in order to get a bolometric flux. 

Once the photometric zero points are determined, \citet{casagrande19} used the colours and \teff~from IRMF to get stellar parameters for their whole sample, adopting and calibrating against the [Fe/H] and log~$g$ for stars in common between SkyMapper and GALAH. Of particular importance to this study is their derivation of photometric metallicities which is aided by the SkyMapper $u$ and $v$ bands. They applied Principal Component Analysis (PCA) on different colour index relations and found that there are typically three components: a primary one due to temperature, and two others that trace [Fe/H] and log $g$. 
They derive the metallicities for 9 million stars using Equation 12 in their paper which relates the metallicity to these colour index relations. To validate their method, they compare these to metallicities in GALAH, APOGEE, and RAVE. Since \citet{casagrande19} used GALAH [Fe/H] for calibration, there is no offset between the two surveys, with residuals having a standard deviation of 0.22 dex. Compared to APOGEE, there is a 0.01 dex offset (APOGEE metallicities being larger) with a 0.25 dex scatter and compared to RAVE, there is a 0.09 dex offset (SkyMapper metallicities being larger) with a scatter of 0.28 dex. Cross-matching the TG sample with the \citet{casagrande19} using the \textit{Gaia} source ID, yields a total of 500,007 stars. We include the \teff~and [Fe/H] from this survey. 

\subsection{Deacon 2019}
\label{sec:deacon}
\citet{deacon19} derived stellar parameters for 939,457 Southern FGK dwarfs that are part of the  \textit{TESS} CTL v7. They use \textit{The Cannon} \citep{ness15} that employs a data-driven approach to parameter estimation. They use stellar colours to infer \teff, log~$g$, [Fe/H], mass, radius, and extinction for the  \textit{TESS} stars. They compiled $u$ and $v$ photometry from SkyMapper survey \citep{wolf18},   $J$, $H$, and $K_S$ from 2MASS \citep{skrutskie06}; $W1$ and $W2$ from WISE \citep{wright10}; as well as \textit{Gaia} $G$ band photometry to build the spectral energy distribution (SED) for the  \textit{TESS} targets. Then, they chose a fourth order function in \teff~ and used the following labels: [Fe/H], log $g$ and \teff~, each label having a separate term and coefficient in the function. They fit this function during the modeling such that it matches the observed colours.  

They chose GALAH as a training set because it is in the Southern hemisphere, similar to the SkyMapper data. The training set does not contain many cool dwarfs ($ T_{\rm eff}$ $<$ 4600 K and log $g$ $>$ 4 dex) or hot objects with $ T_{\rm eff} > 7000$ K. With their parameter estimation, they derived \teff~and log~$g$ for 939,457 stars while a subset of that, containing 638,972 stars, has [Fe/H] estimates. 


We cross-match this catalogue with the TG sample using the  \textit{TESS} ID and apply the following cuts as prescribed by the authors: 1) parameter $\neq$ -99.9 (i.e. outside the training set bounds) 2) exclude stars hotter than 7000 K, giant stars cooler than 4000 K, and dwarfs cooler than 4600K are excluded, 3) stars that do not have [Fe/H] entries are removed, and 4) stars with [Fe/H]$>$0.55 dex, which have bad estimates and are therefore flagged in their catalogue, are removed. This leaves us with 413,100 cross-matches with the TG sample. We use [Fe/H], log~$g$, and $T_{\rm eff}$ from this catalogue.

\section{Results}

\label{sec:results}

\subsection{Kinematics}
\label{sec:kinematics}

\begin{figure*}
\includegraphics[width=0.45\textwidth]{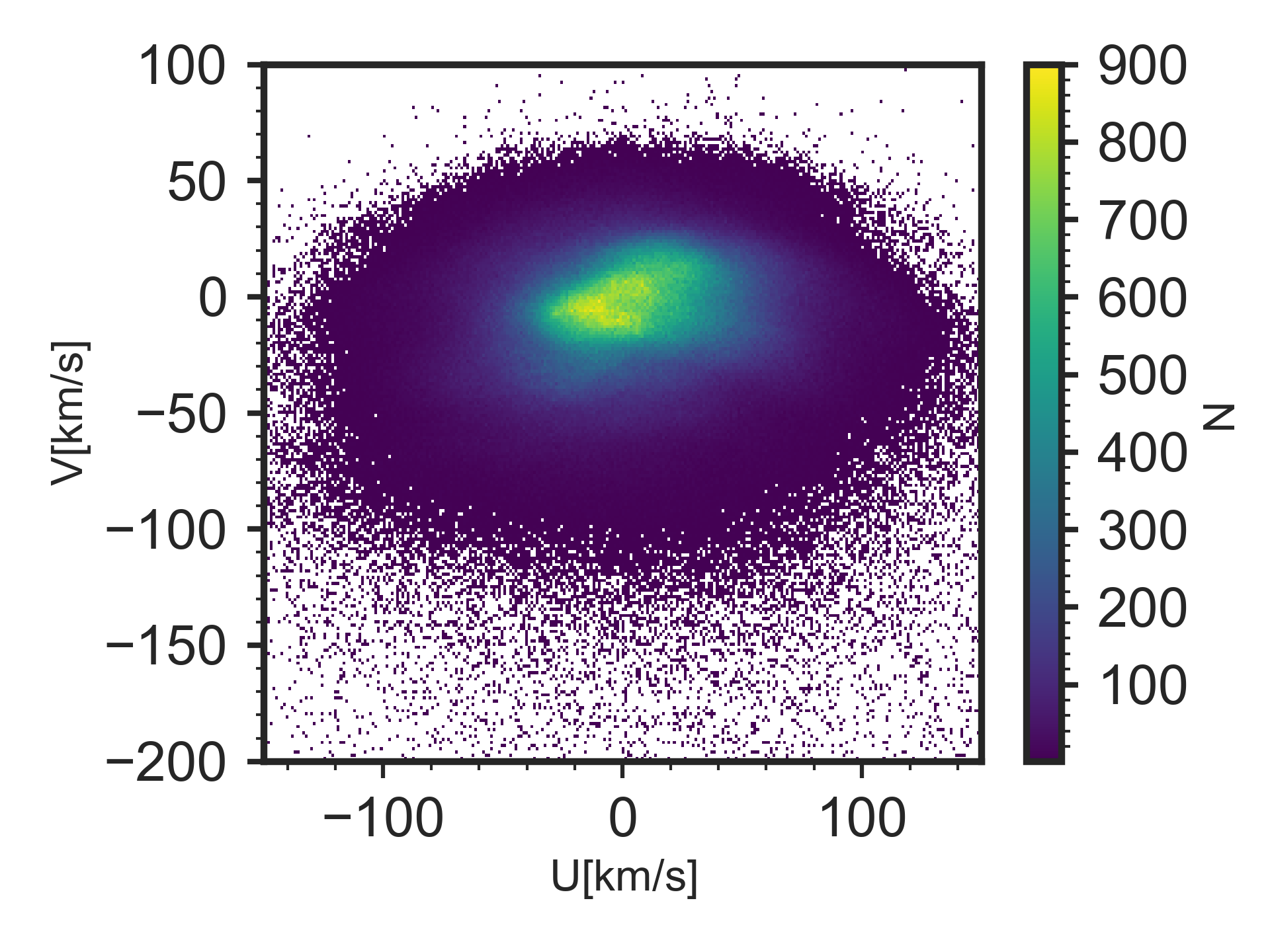}
\includegraphics[width=0.45\textwidth]{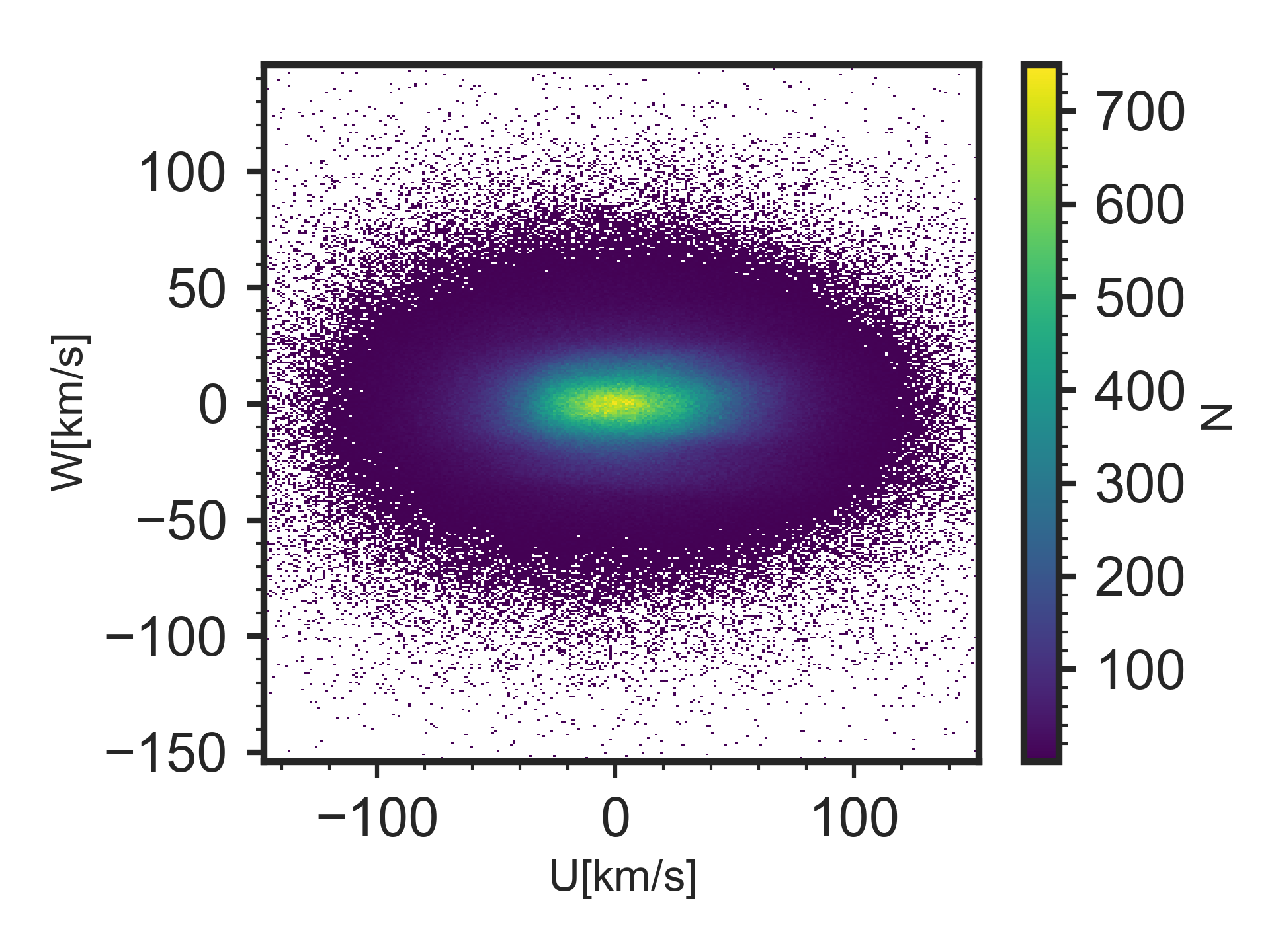}
\includegraphics[width=0.45\textwidth]{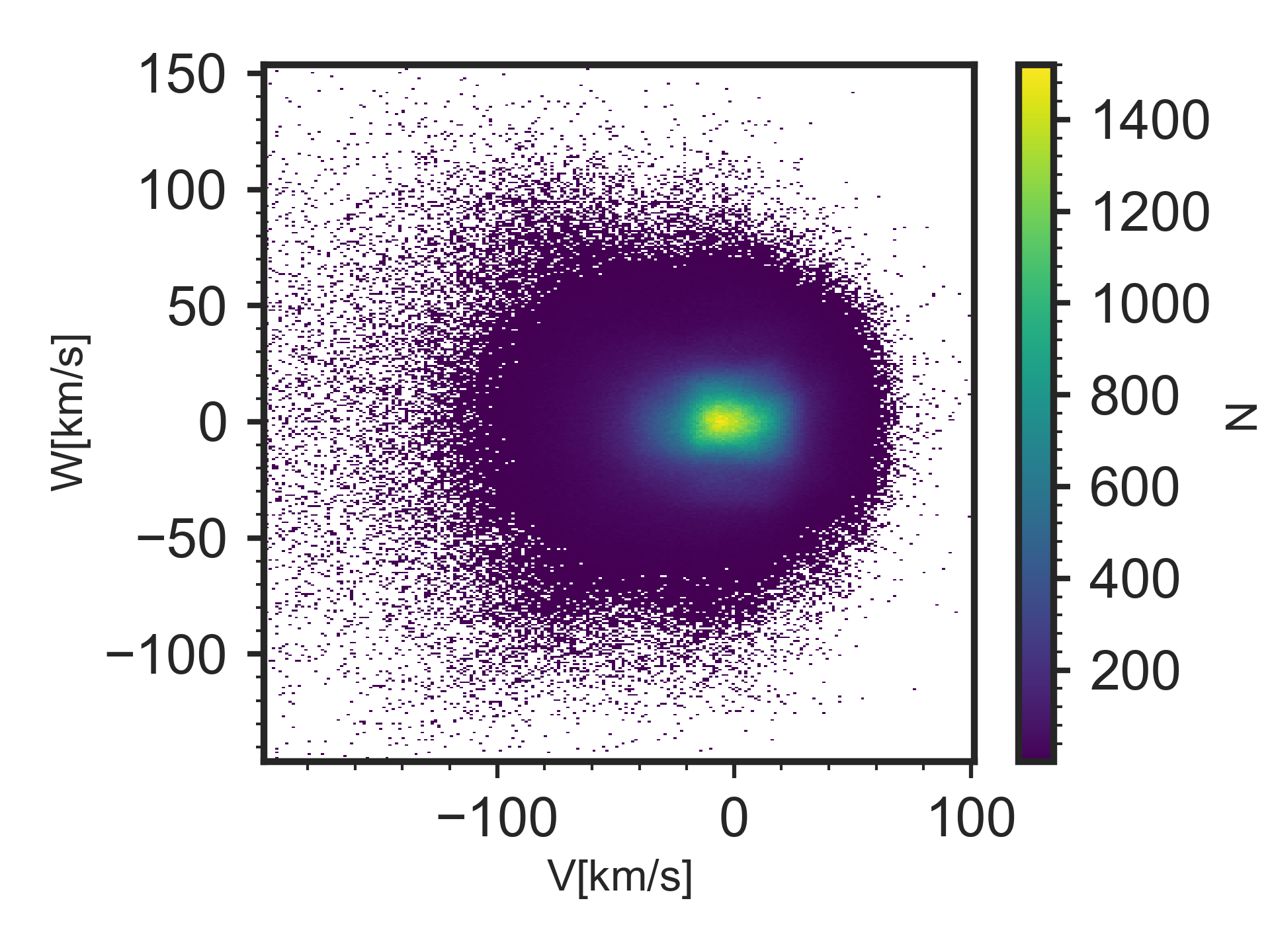}
\caption{2D distributions of the UV (left), UW (center), and VW (right) kinematics in LSR with the colourbar indicating the density of stars. The UV plane shows substructures attributed to moving groups \citep{antoja08,bovy09,gaiakinematics}. On the other hand, the UW and VW kinematics show more isotropic distributions.} 
\label{fig:UV}
\end{figure*}

\begin{figure}
\center
\includegraphics[width=0.4\textwidth]{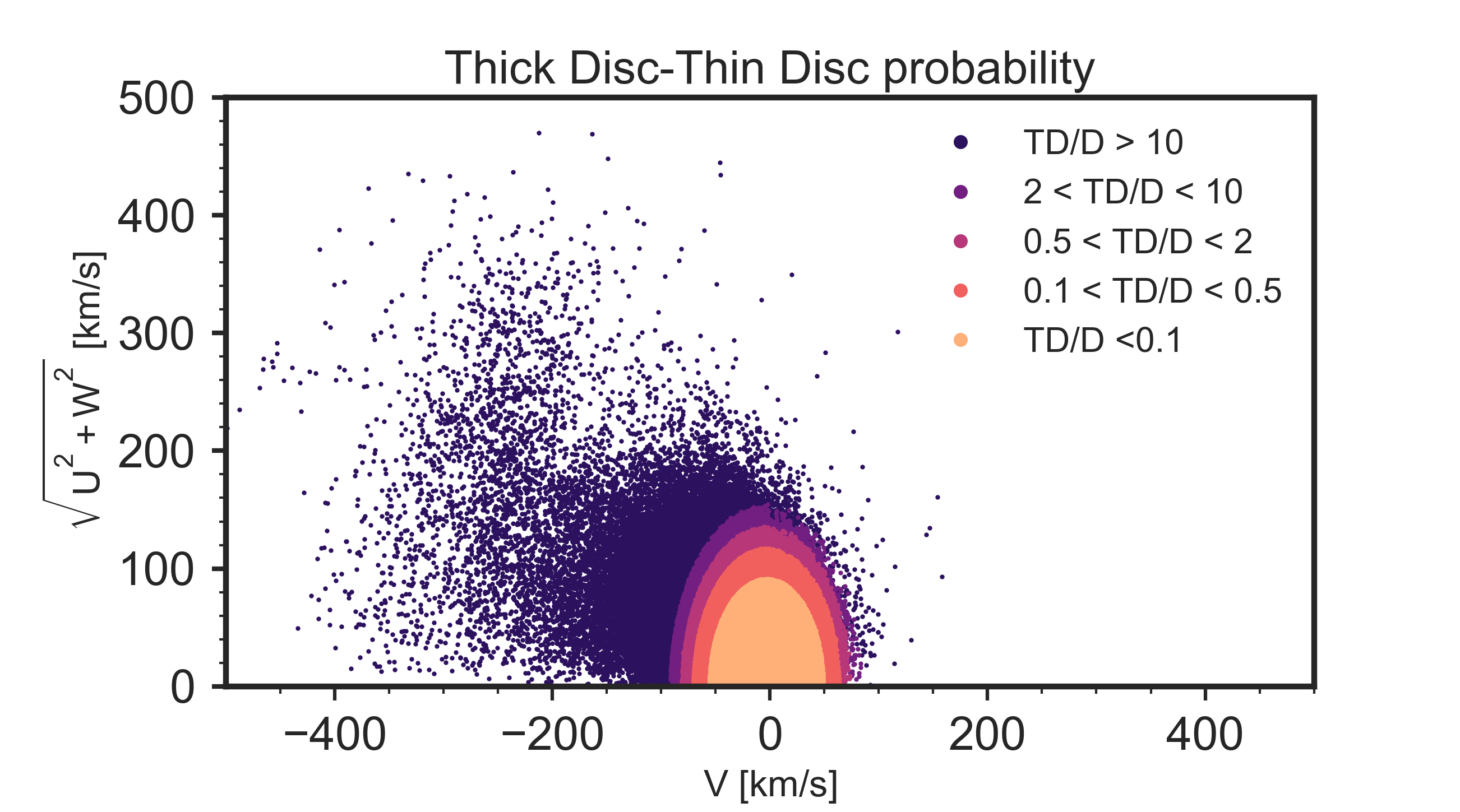}
\includegraphics[width=0.4\textwidth]{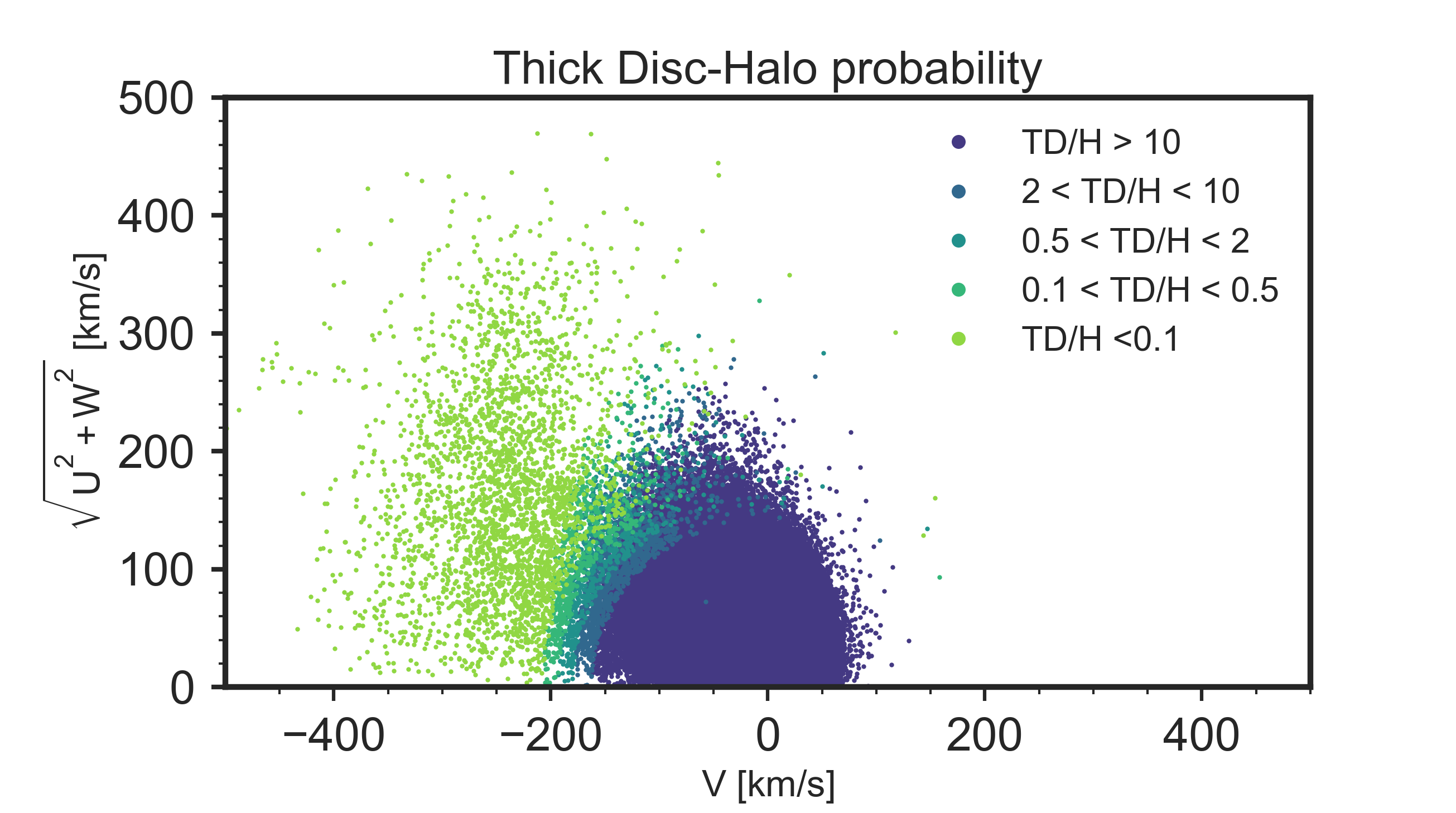}
\caption{Top panel: Toomre diagram for the TG sample colour-coded by the thick disc to thin disc probability, where purple ($TD/D > 10$) denotes that a star is 10 times more likely to be part of the thick disc than the thin disc and orange ($TD/D < 0.1$) indicates a star is 10 times more likely to be part of the thin disc than the thick disc. Bottom panel: Toomre diagram for the TG sample colour-coded by the thick disc to halo membership where green ($TD/H < 0.1$) represents a star is 10 times more likely to be part of the halo than the thick disc and vice versa.} 
\label{fig:toomre}
\end{figure}

We adopt the kinematics from \citet{marchetti18} that use $Gaia$ DR2 radial velocities, parallaxes, and proper motions. We refer the reader to their paper for details on deriving full 6D phase space information  but we briefly discuss it here. One needs the distance to the source, derived from the parallax, $\varpi$, in order to convert an apparent motion in the sky to an actual velocity with respect to the Galactic center. This total velocity is calculated using two approaches based on the relative parallax error, $ f = \sigma_{\varpi} / \varpi $. For the first approach, the sample with the low parallax errors i.e. $0 < f < 0.1$, has distances derived from simply inverting the parallax. For the second approach, the sample with higher parallax errors i.e. $f > 0.1$ has distances derived from a Bayesian analysis following \citet{bailerjones15}, where they use a weak distance prior (i.e. exponentially decreasing space density prior) that changes with Galactic latitude and longitude. 

The final data products include Galactic Cartesian velocities (U,V,W) which we convert to Local Standard of Rest (LSR) by subtracting $\rm V = 238~km s^{-1}$, the adopted rotation velocity at the position of the Sun from \citet{marchetti18}. Figure \ref{fig:UV} shows the velocities in the UV, VW, and UW reference planes. We follow the convention where U is positive towards the direction of the Galactic center (GC), V is positive for a star with the same rotational direction as the Sun going around the galaxy, with 0 at the same rotation as sources at the Sun's distance, and $W$ positive towards the north Galactic pole. Particularly in the UV plane, one can see substructures because of the presence of moving groups that have similar kinematics, possibly due to resonances with the bar and spiral arms \citep{skuljan99,quillen05,antoja08,bovy09,gaiakinematics,trick19}. 

We provide a quantitative metric for the the thin disc, thick disc, and halo membership of \textit{TESS} targets,  following the membership determination from Appendix A of \citealt{bensby14}. There are other, more robust ways of determining the Galactic component membership of stars through using ages or [$\alpha$/Fe] abundance; however, kinematics provide the simplest criteria that could be applied to the majority of the $TESS$ targets. We therefore adopt this method, but caution the reader about the contamination between each Galactic components as would be discussed in Section 4. 

This method assumes that the space velocities have Gaussian distributions,
\begin{equation}
f = k \cdot exp \Bigg( \frac{(U_{LSR} - U_{asym})^2}{2 \sigma^2_U} - \frac{(V_{LSR} - V_{asym})^2}{2 \sigma^2_V} - \frac{W_{LSR}^2}{2 \sigma^2_W} \Bigg)
\end{equation}
with k defined as $((2\pi)^{3/2}\sigma_U \sigma_V \sigma_W)^{-1}$ for normalization, $\sigma_U$,$\sigma_V$, and $\sigma_W$ are the velocity dispersions for each Galactic component, and $V_{asym}$ and $U_{asym}$ are the asymmetric drifts i.e. the mean tangential speed deviations from a circular velocity. We use the values listed in Table \ref{tab:bensby} to establish the relative probability for thick disc (TD) to thin disc (D) membership:
\begin{equation}
\frac{TD}{D} = \frac{X_{TD}}{X_{D}} \cdot \frac{f_{TD}}{f_D}
\end{equation}
where $X$ is defined as the observed fraction of the stellar population in the Solar neighborhood. Thick disc to halo membership is calculated the same way. The majority of the TG sample as well as the majority of the sample from \citet{bensby14} are at distances well within the scale length of the thin disc. It is therefore adequate to assume the Solar neighborhood stellar population fractions for the rest of our sample. 

\begin{table}
    \centering
    \begin{tabular}{c|c}
        \hline
        \hline
        Component & $X$ \\
        \hline
        Thin disc (D) & 0.85 \\
        Thick Disc (TD) & 0.09\\
        Halo (H) & 0.0015\\
        \hline
        \hline
    \end{tabular}
    \caption{Observed fraction of stellar population in the solar neighborhood, adopted from Table A.1 from \citet{bensby14}.}
    \label{tab:bensby}
\end{table}

We estimate the errors in the membership by performing a series of 500 Monte Carlo simulations, perturbing the observed Cartesian space velocities U,V, and W, by the errors reported in \citet{marchetti18} and adopting a Gaussian distribution for these errors.  
This leads to a distribution in $f$ for the thin disc, thick disc, and halo, and consequently, a distribution in their relative probabilities. We quote the median values for $TD/D$ and $TD/H$ and the 16th and 84th percentile for the lower and upper bounds, respectively, as not all the $f$ distributions are Gaussian in shape. 

We show Toomre diagrams in Figure \ref{fig:toomre}, with the vertical axis calculated as $\sqrt{U^2 + W^2}$. The diagrams are colour-coded by relative membership to the thin disc, thick disc, and halo. The Toomre diagram is helpful in determining the structure-membership because of the different contributions of rotation and velocity dispersion to each stellar population. In general, the disc shows evidence of rotation seen in the symmetry in the range of velocities around $\rm V_{LSR}$ = 0 $\rm kms^{-1}$, i.e. $\rm V = 238 km s^{-1}$, while the halo shows evidence of pressure support from high velocity dispersion. Using the relative probabilities, we assign the component memberships accordingly: 

\begin{itemize}
    \item{thin disc: $TD/D < 0.5$ and $TD/H > 2$}
    \item{thick disc: $TD/D > 2$ and $TD/D > 2$}
    \item{halo: $TD/H < 0.5$ and $TD/D > 2$}    
    \item{thick disc/thin disc: $0.5 < TD/D < 2$ and $TD/H > 2$}
    \item{thick disc/halo: $0.5 < TD/H < 2$ and $TD/D > 2$}.    
\end{itemize}

We find that 91.77\% of the stars in the TG sample are in the thin disc, 3.64\% in the thick disc, 0.16\% in the halo, 3.00\% in the thick disc - thin disc transition region, and 0.03\% in the thick disc - halo transition region. The majority of the stars in the TG sample are from the kinematically defined thin disc. This is expected because \textit{TESS} is targeting the nearest and brightest stars which would naturally originate from the thin disc. Further, the adopted values for the fraction of stellar population gives $X_{D}/X_{TD}$ = 9.4, that is, we are 9 times more likely to find thin disc stars than thick disc stars. For comparison, we have applied other stellar population fractions in the literature (e.g. \citealt{reddy05} with 0.93, 0.7, and 0.006 for the thin disc, thick disc, and halo, respectively) to get the kinematic membership of stars. This yielded similar contributions from each Galactic component at 94.16\% for the thin disc, 3.14\% for the thick disc, and 0.20\% for the halo.

\subsection{Atmospheric Parameters and Chemistry}
\label{sec:chemistry}

\begin{figure}
\center
\includegraphics[width=0.45\textwidth]{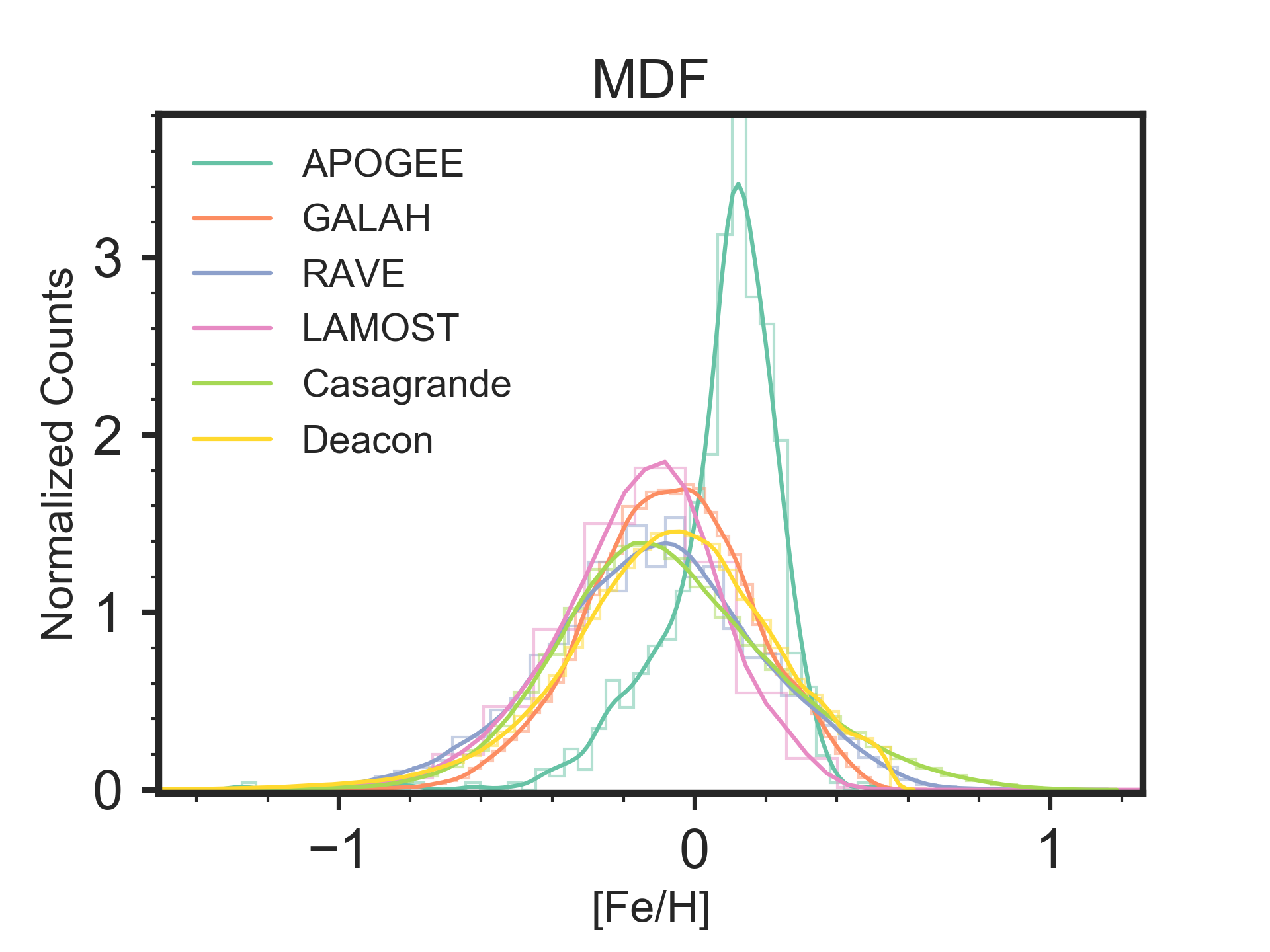}
\caption{MDF of the TG sample cross-matched with APOGEE (teal), GALAH (orange), RAVE (purple), LAMOST (pink), \citealt{casagrande19} (green), and \citealt{deacon19} (yellow). We note that these MDFs peak at solar and higher metallicities, consistent with being dominated by the thin disc.}
\label{fig:mdf}
\end{figure}

\begin{figure*}
\includegraphics[width=0.95\textwidth]{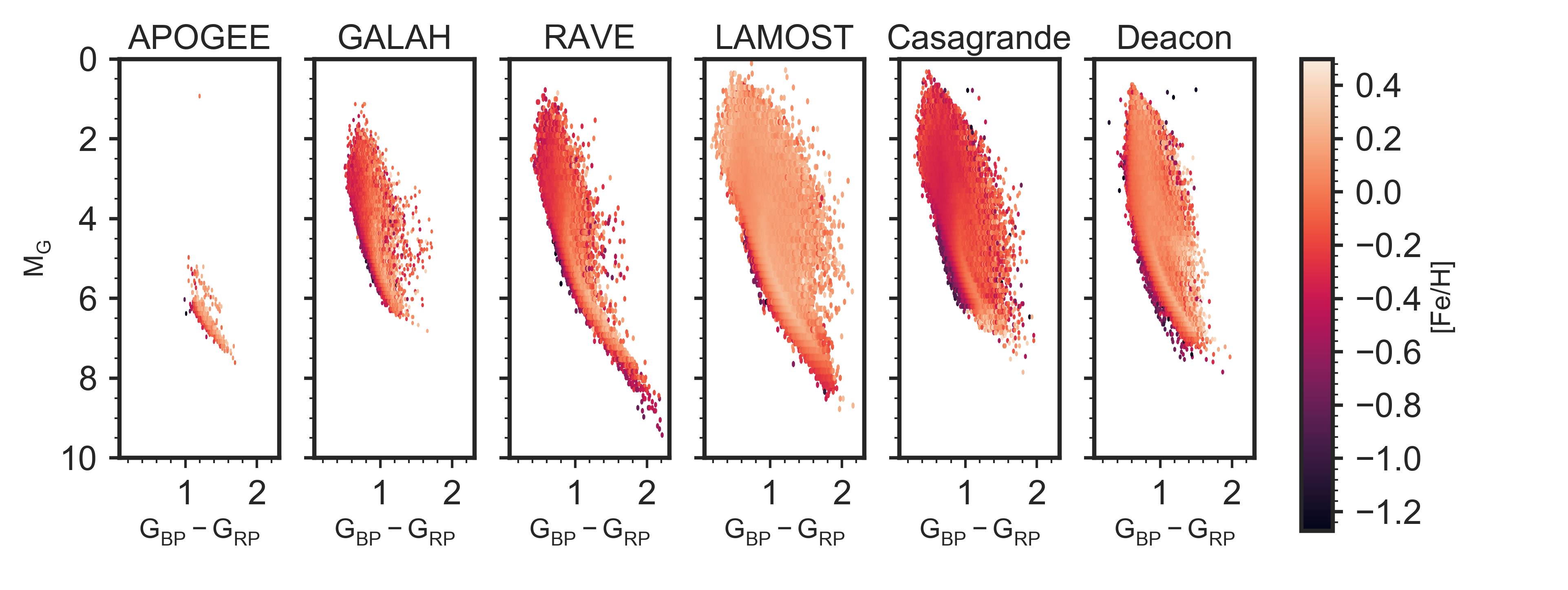}
\caption{From left to right: Gaia G-band magnitude vs $G_{BP} - G_{RP}$ colours for stars in common between the TG sample and APOGEE, GALAH, RAVE, LAMOST, \citet{casagrande19}, and \citet{deacon19}, colour-coded by [Fe/H]. These color magnitude diagrams show that the majority of the TG-crossmatched samples are dwarfs.}
\label{fig:loggteff}
\end{figure*}

We construct metallicity distribution functions (MDF) of the cross-matched TG sources with APOGEE, RAVE, GALAH, LAMOST, \citet{casagrande19}, and \citet{deacon19} as shown in Figure \ref{fig:mdf} with the histogram normalized due to the varying number of targets available from each survey. We also show the CMD for these surveys in Figure \ref{fig:loggteff} colour-coded by the metallicity. It is apparent in each of the survey CMD that the cross-matched TESS targets are indeed dwarfs and for the most part are the stars with lower absolute magnitudes (i.e. brighter than $\rm M_G \sim$7.5 mag) in the bigger TG sample (see Fig. \ref{fig:cmd}), barring RAVE and LAMOST. The survey that shows the most difference is the APOGEE-crossmatched sample which covers a smaller parameter space than the other surveys, and contains stars with lower intrinsic brightness. 

\begin{figure}
\center
\includegraphics[width=0.35\textwidth]{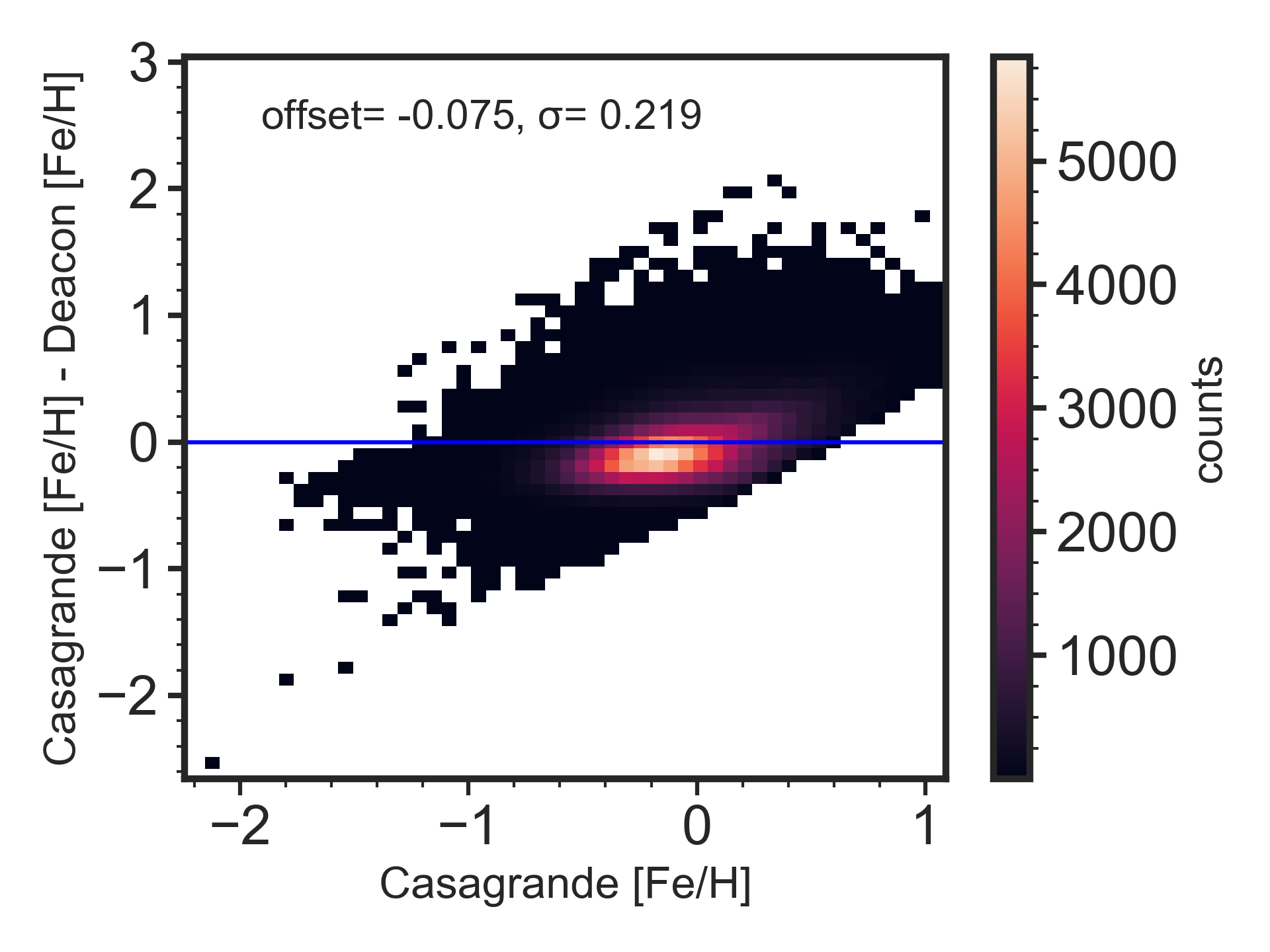}
\caption{Difference between [Fe/H] from \citet{casagrande19} and from \citet{deacon19} vs [Fe/H] from \citet{casagrande19}. The horizontal line shows where there is no difference between the two photometric metallicity catalogues. In general, the [Fe/H] from \citet{casagrande19} is lower than those from \citet{deacon19}.}
\label{fig:DC}
\end{figure}

We obtained the difference in metallicities between each survey and find typical offsets of 0.00 to 0.50 dex with a median of 0.04 dex and scatters in the differences between 0.08 dex and 0.31 dex, with a median of 0.22 dex.  Here we highlight the comparison between 249,152 stars in common between \citet{casagrande19} and \citet{deacon19} as shown Figure \ref{fig:DC}, which are both photometrically-derived. There is a positive trend in the difference between the two catalogues as well as a systematic offset of $\Delta$[Fe/H] = -0.08 $\pm 0.22$ dex, with \citet{casagrande19} being lower. Even though both surveys use SkyMapper photometry, the methods to derive [Fe/H] are different for both studies as discussed in Sections \ref{sec:casagrande} and \ref{sec:deacon}, respectively. Though the offset between  \citet{casagrande19} and \citet{deacon19} is higher compared to the median offset across the whole sample, the scatter in the difference is similar to the median trend for the whole sample. 

\begin{figure*}
\includegraphics[width=0.9\textwidth]{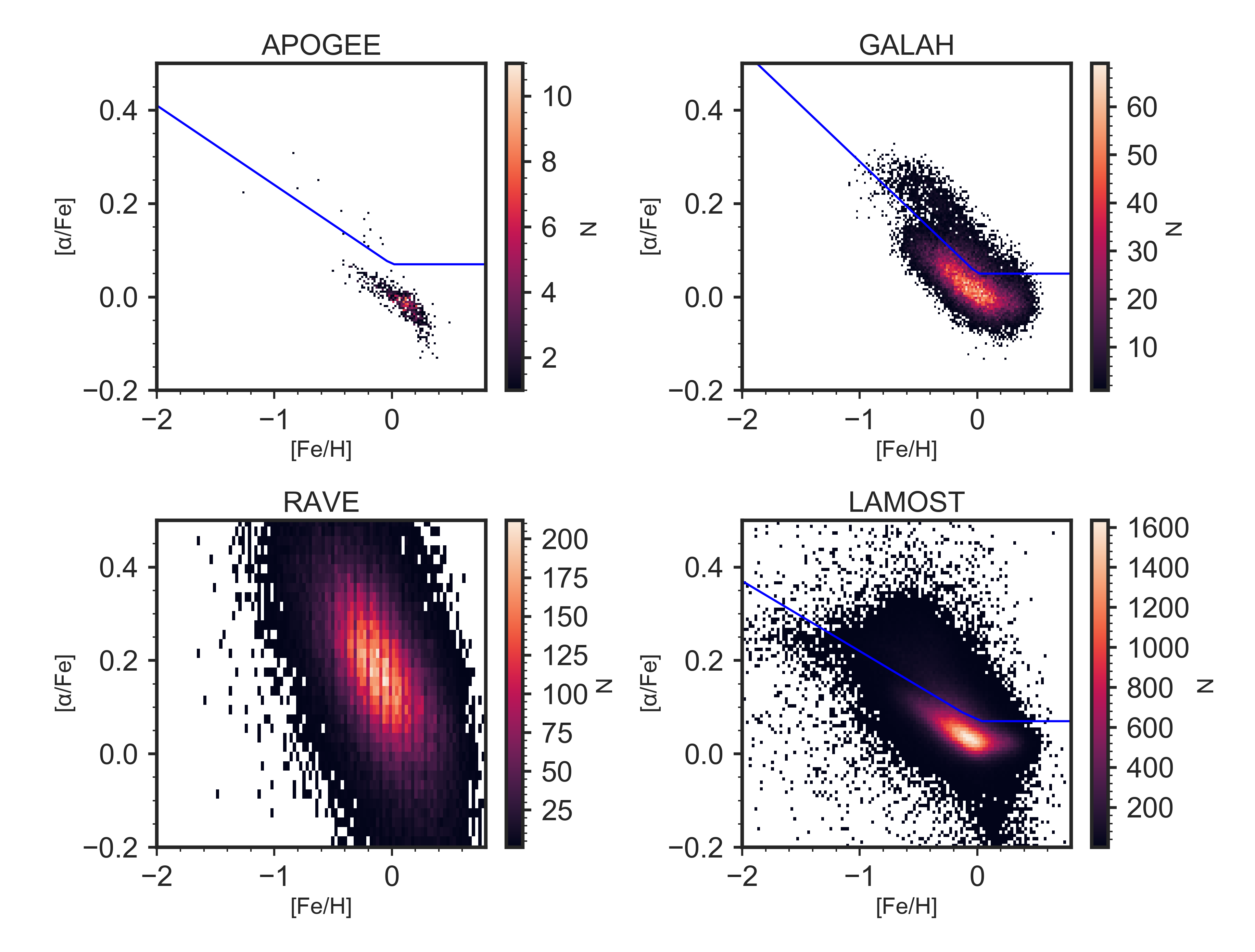}
\caption{Left to right from top to bottom: 2D histograms of [$\alpha$/Fe] vs [Fe/H] for the TG sample cross-matched with APOGEE, GALAH, RAVE, and LAMOST. Orange to black illustrates higher density to lower density of sources. GALAH shows two distinct trends, a high-$\alpha$ stellar population and a low-$\alpha$ stellar population. This trend is weaker in TG-APOGEE because of a smaller sample size in the high-$\alpha$ track while this trend is indistinguishable in TG-RAVE. We mark the separation between the low-$\alpha$ and high-$\alpha$ stellar populations in APOGEE, GALAH, and LAMOST to guide the eye. }
\label{fig:alphaFe}
\end{figure*}

Next, we explore the [$\alpha$/Fe] of the TG host stars that are also in APOGEE, GALAH, RAVE, and LAMOST shown in Figure \ref{fig:alphaFe}. 
[$\alpha$/Fe] is a powerful metric of star formation history, with high [$\alpha$/Fe] indicating rapid star formation. 
The [$\alpha$/Fe] vs [Fe/H] diagram is useful in distinguishing the chemical tracks for different stellar populations in the Milky Way (e.g., \citealt{edvardsson93,adibekyan11, haywood13,nidever14,hawkins15,hayden15, buder19}). Thick disc and halo stars have higher [$\alpha$/Fe] and lower [Fe/H] compared to the thin disc, which shows solar [$\alpha$/Fe] and higher metallicities. APOGEE, GALAH, and LAMOST show these trends with the two tracks, although the low metallicity, high-$\alpha$ track is less populated, especially for APOGEE.  The TG sample cross-matched with GALAH has an [$\alpha$/Fe] vs [Fe/H] trend that agrees with the larger GALAH sample \citep{buder19}. This makes sense as both the TESS CTL and GALAH overlap in their target sample i.e. nearby stars. On the other hand, the TG sample crossmatched with APOGEE has no giants by design, therefore the [alpha/Fe] vs [Fe/H] trend in our study does not reflect that of the larger APOGEE sample \citep{holtzman18}. The RAVE [$\alpha$/Fe] vs [Fe/H] track does not show the same dichotomy as the other surveys. This is due to RAVE having lower resolution (R$\sim$7500) as well as a significantly shorter wavelength range (8410-8795 \AA) compared to APOGEE and GALAH. The uncertainty in the [Fe/H] and [alpha/Fe] in RAVE are both ~0.2 dex, which also affects the distinction between the thin disc and thick disc trends (see Section 8 in \citealt{kunder17}). Nonetheless, we decide to report the [$\alpha$/Fe] from RAVE as we have applied the necessary flags suggested by \citet{kunder17}. 

\subsection{The Catalogue}
\label{sec:thecatalog}
The chemo-kinematic properties of  \textit{TESS} host stars from the CTL cross-matched with $Gaia$ DR2, APOGEE, GALAH, LAMOST, RAVE, \citet{casagrande19}, and \citet{deacon19} are provided as a table with the columns listed in Table \ref{tab:catalogue}. The catalogue consists of the TG sample and includes information from the \textit{TESS} CTL, astrometry and photometry from $Gaia$ DR2, 6D phase space information from \citet{marchetti18}, distances from \citet{bailerjones18}, \teff, log~$g$, spectroscopic and photometric [Fe/H], and [$\alpha$/Fe] from APOGEE, GALAH, RAVE, LAMOST, \citet{casagrande19} and \citet{deacon19} where available, and kinematic membership probabilities to the thin disc, thick disc, and halo (as discussed is Section \ref{sec:kinematics}). For ease of use, we also provide absolute membership to the thin disc, thick disc, and halo, by imposing $D + TD + H =1$, and deriving $D$, $TD$, and $H$ from this relation. We refer to the whole catalogue as TGv8.  

\section{Discussion}
\label{sec:discussion}

\subsection{Tying Chemistry and Kinematics}
\label{sec:chemo-kinematics}

\begin{figure*}
\includegraphics[width=0.45\textwidth]{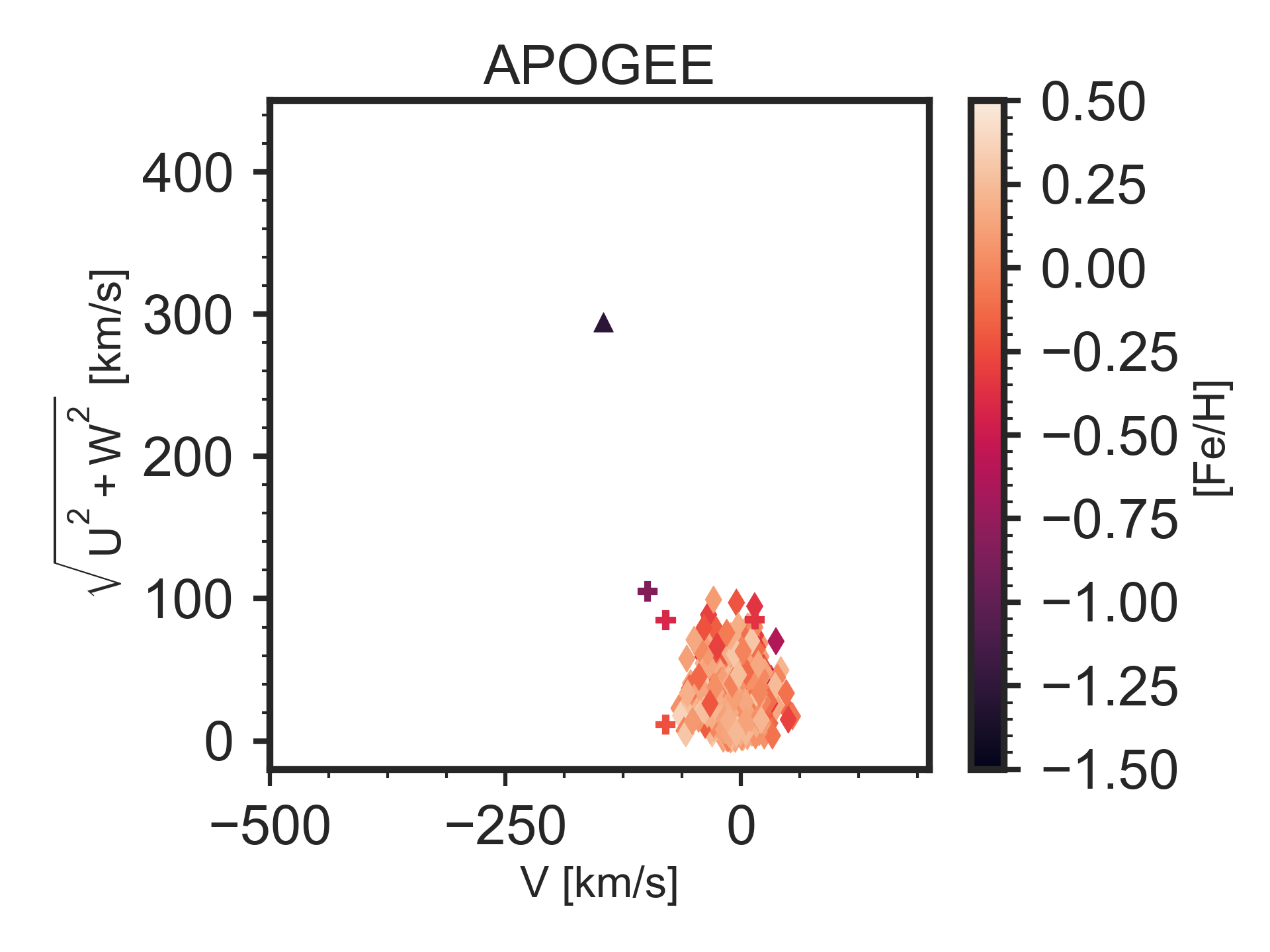}
\includegraphics[width=0.45\textwidth]{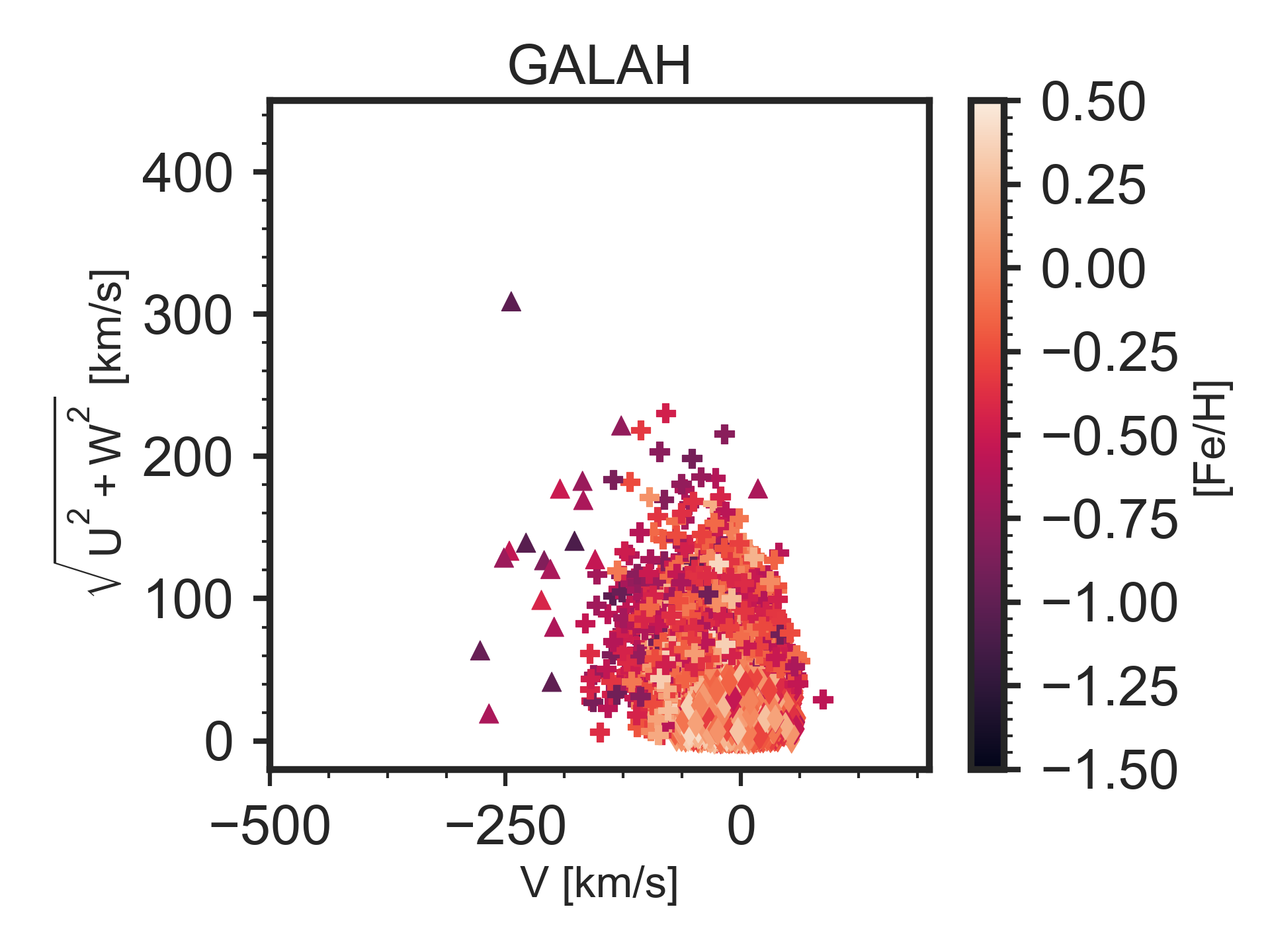}
\includegraphics[width=0.45\textwidth]{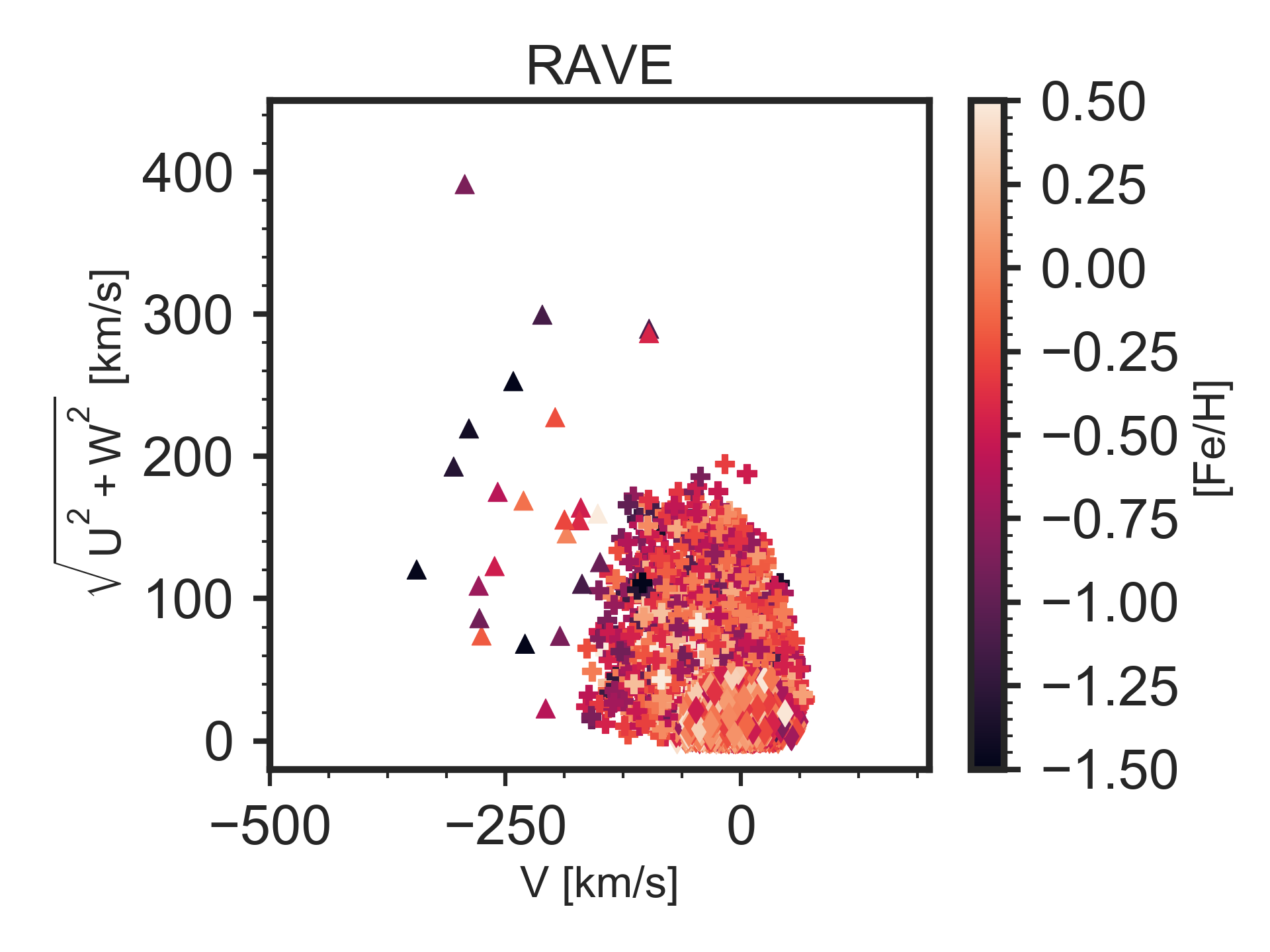}
\includegraphics[width=0.45\textwidth]{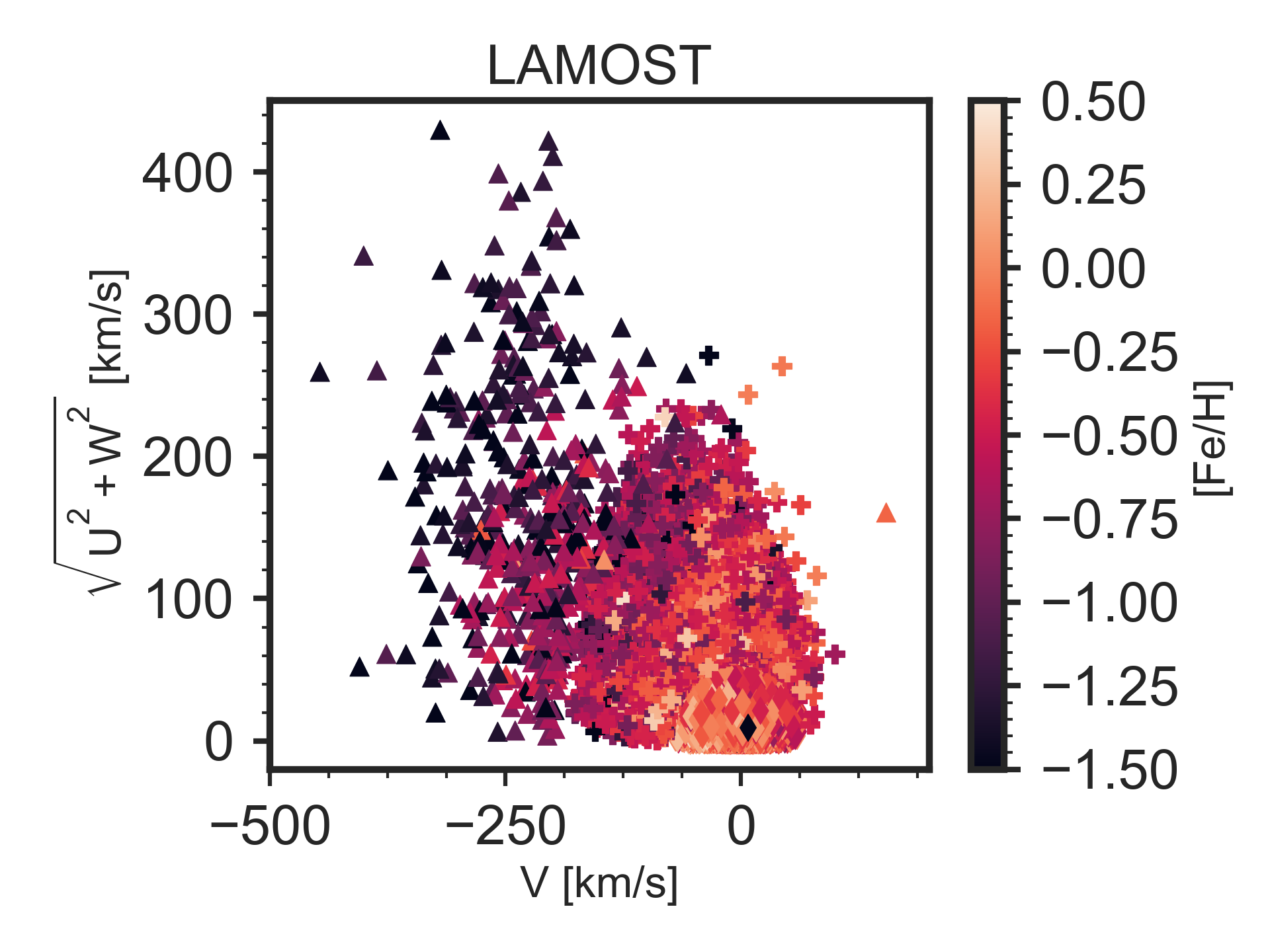}
\includegraphics[width=0.45\textwidth]{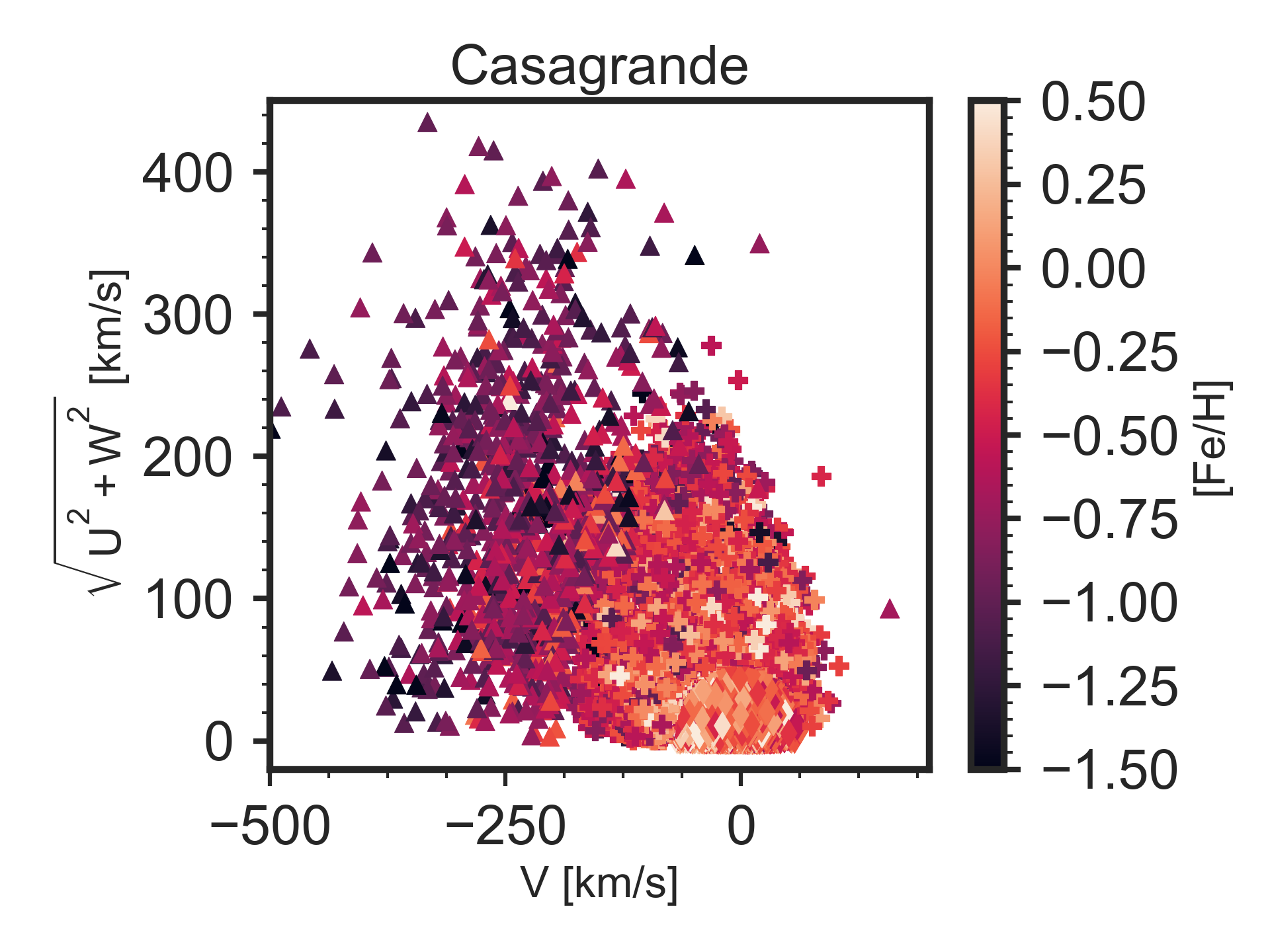}
\includegraphics[width=0.45\textwidth]{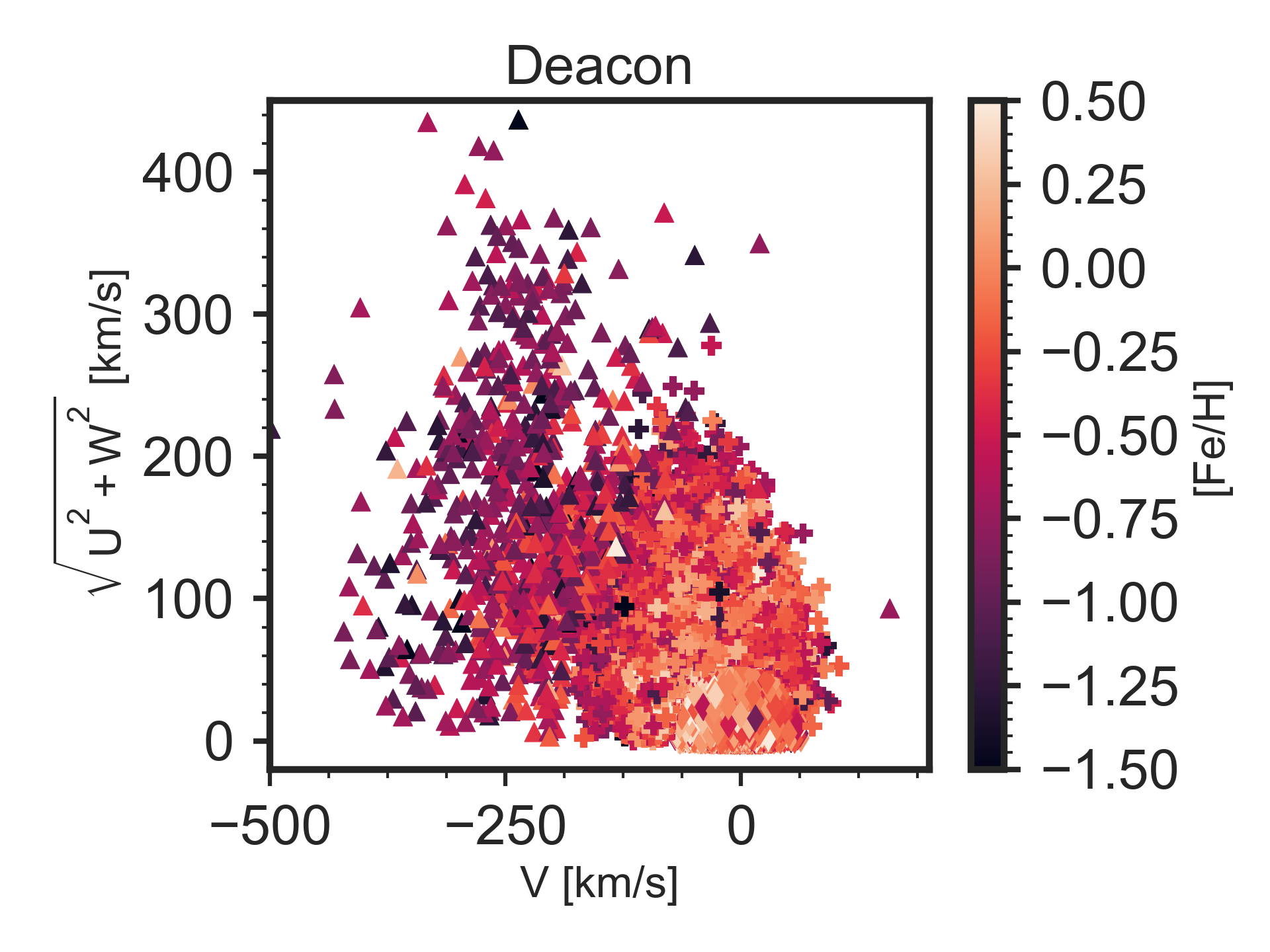}
\caption{Toomre diagram (in LSR) colour-coded by metallicity for the TG sample cross-matched with APOGEE, GALAH, RAVE, LAMOST, \citealt{casagrande19} and \citealt{deacon19}. We also show different symbols for the kinematic thin disc (diamond), thick disc (plus), and halo (triangle) stars as discussed in Section \ref{sec:kinematics}.}
\label{fig:kin_chem_FeH_spec}
\end{figure*}

\begin{figure*}
\includegraphics[width=0.45\textwidth]{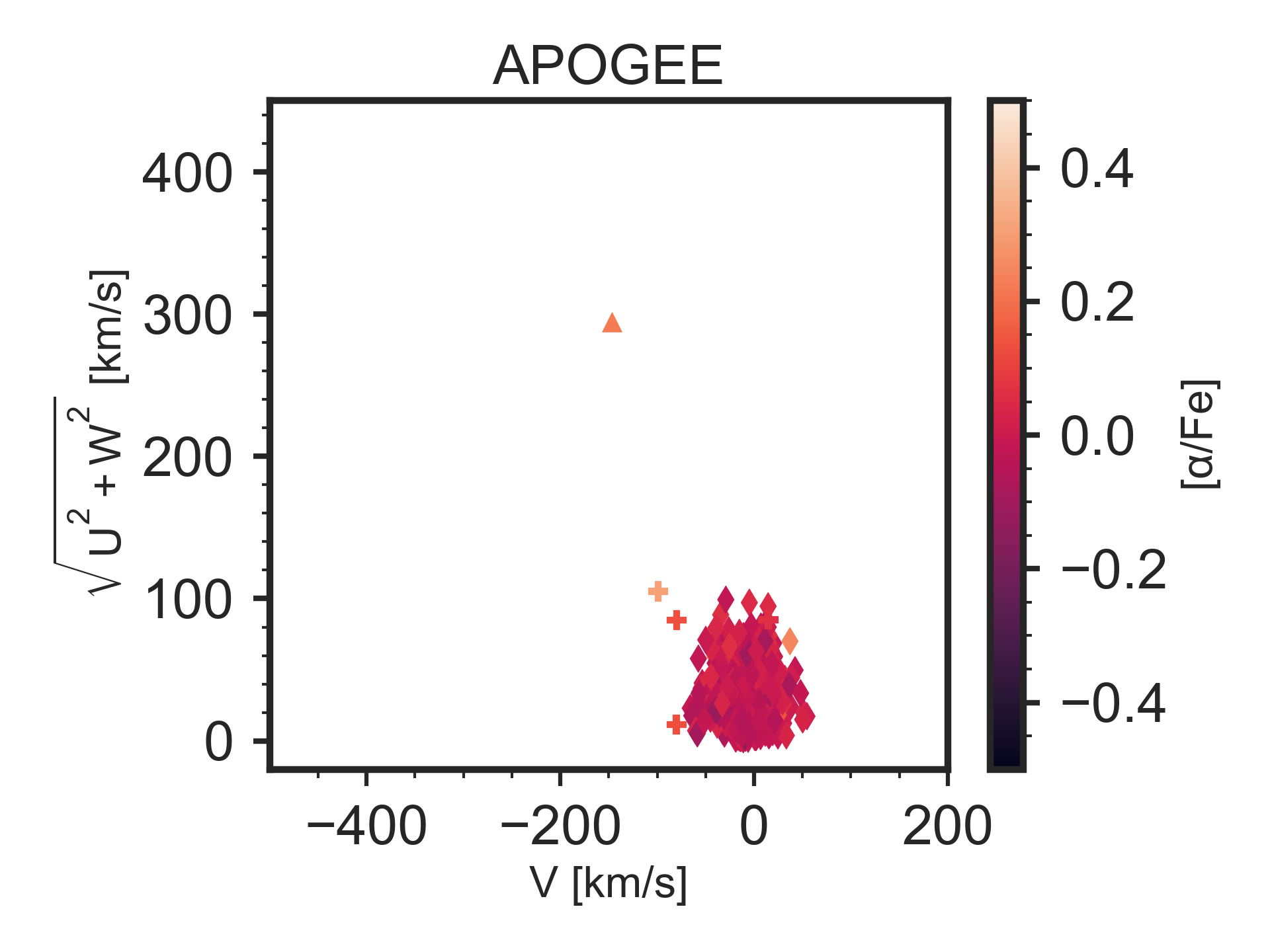}
\includegraphics[width=0.45\textwidth]{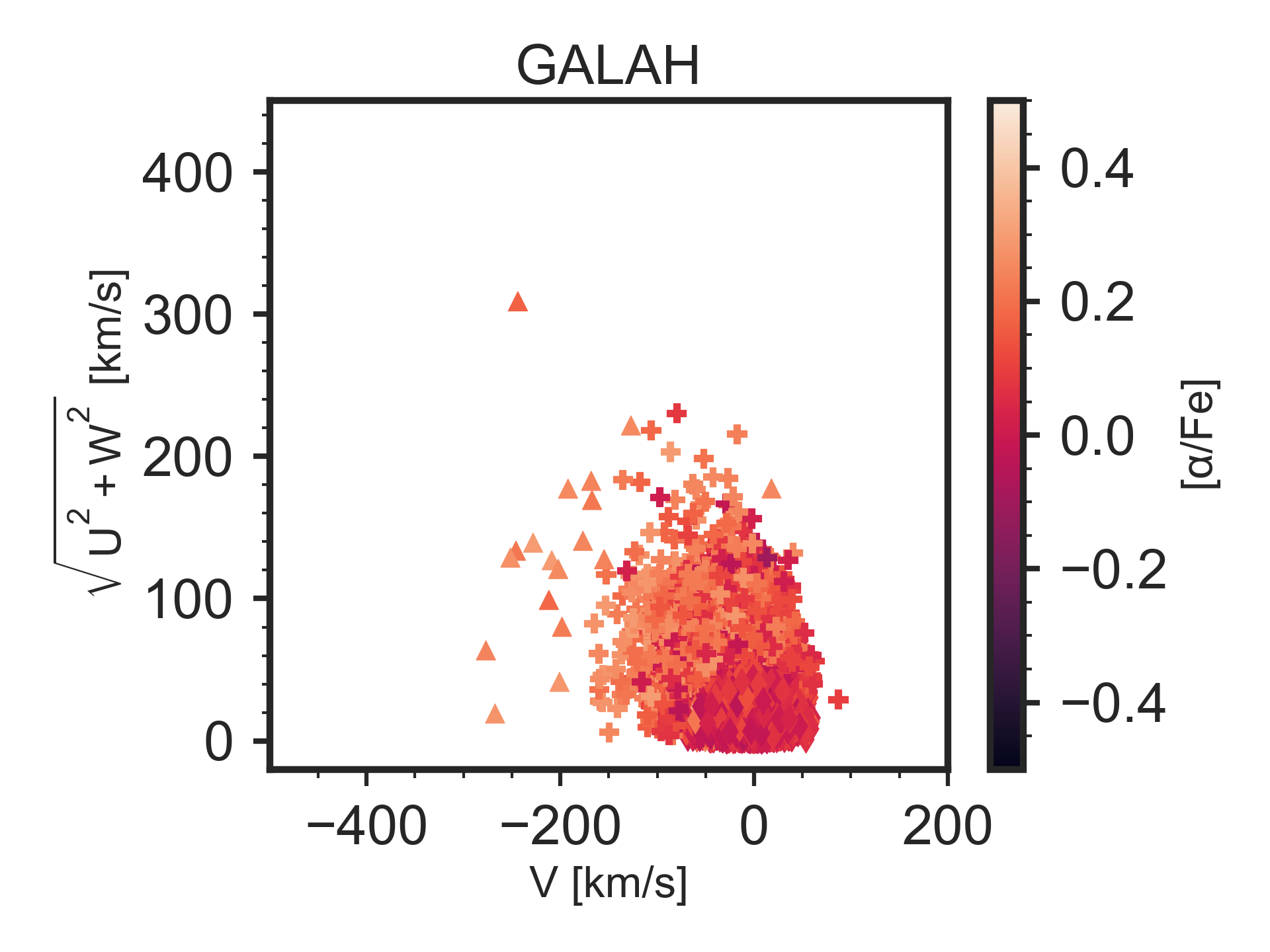}
\includegraphics[width=0.45\textwidth]{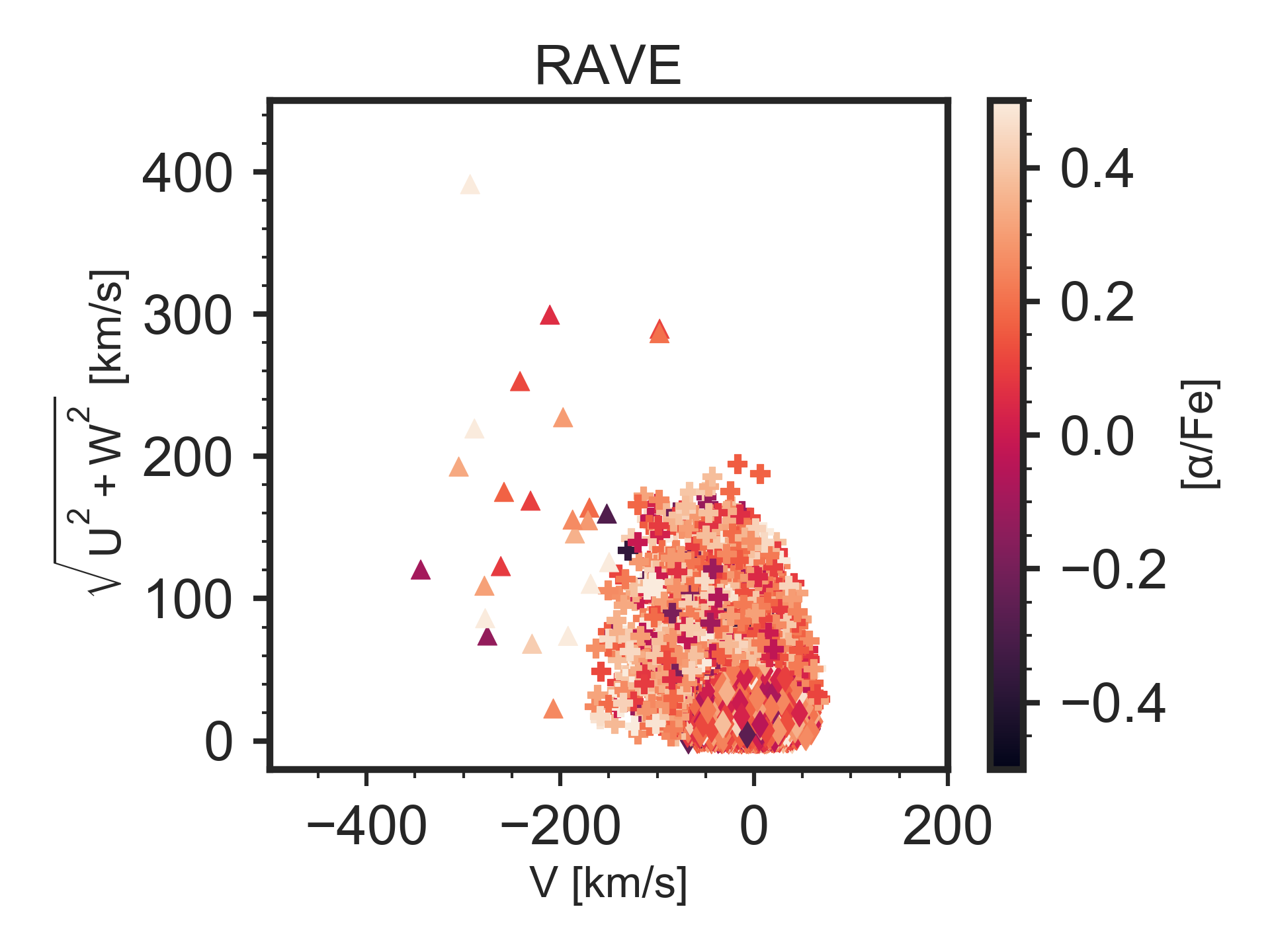}
\includegraphics[width=0.45\textwidth]{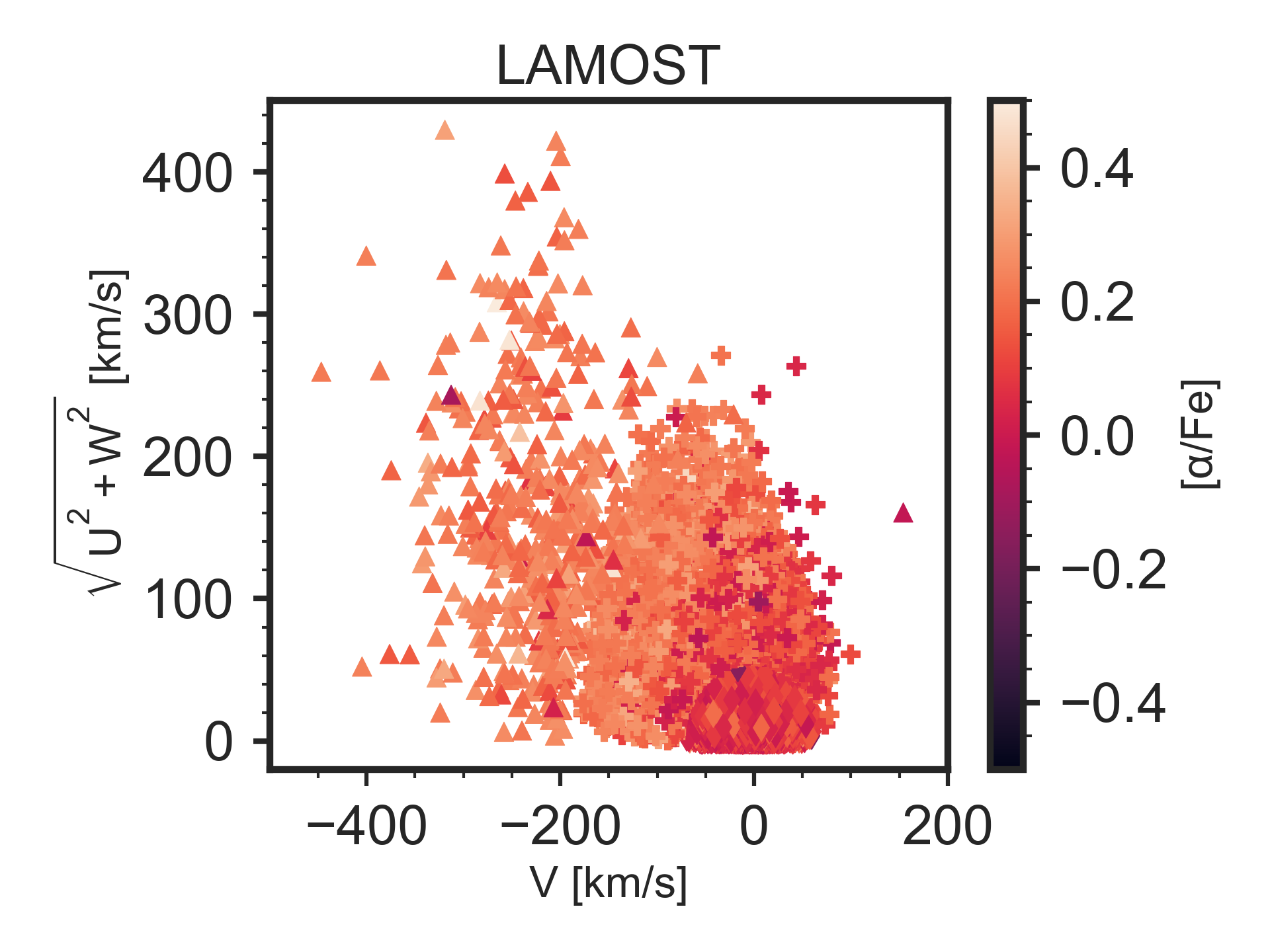}
\caption{Left to right from top to bottom: Toomre diagram (in LSR) colour-coded by [$\alpha$/Fe] abundance ratio for the TG sample cross-matched with APOGEE, GALAH, RAVE, and LAMOST with different symbols for the kinematic thin disc (diamond), thick disc (plus), and halo (triangle) stars.}
\label{fig:kin_chem_alpha}
\end{figure*}

\begin{figure*}
\includegraphics[width=0.60\textwidth]{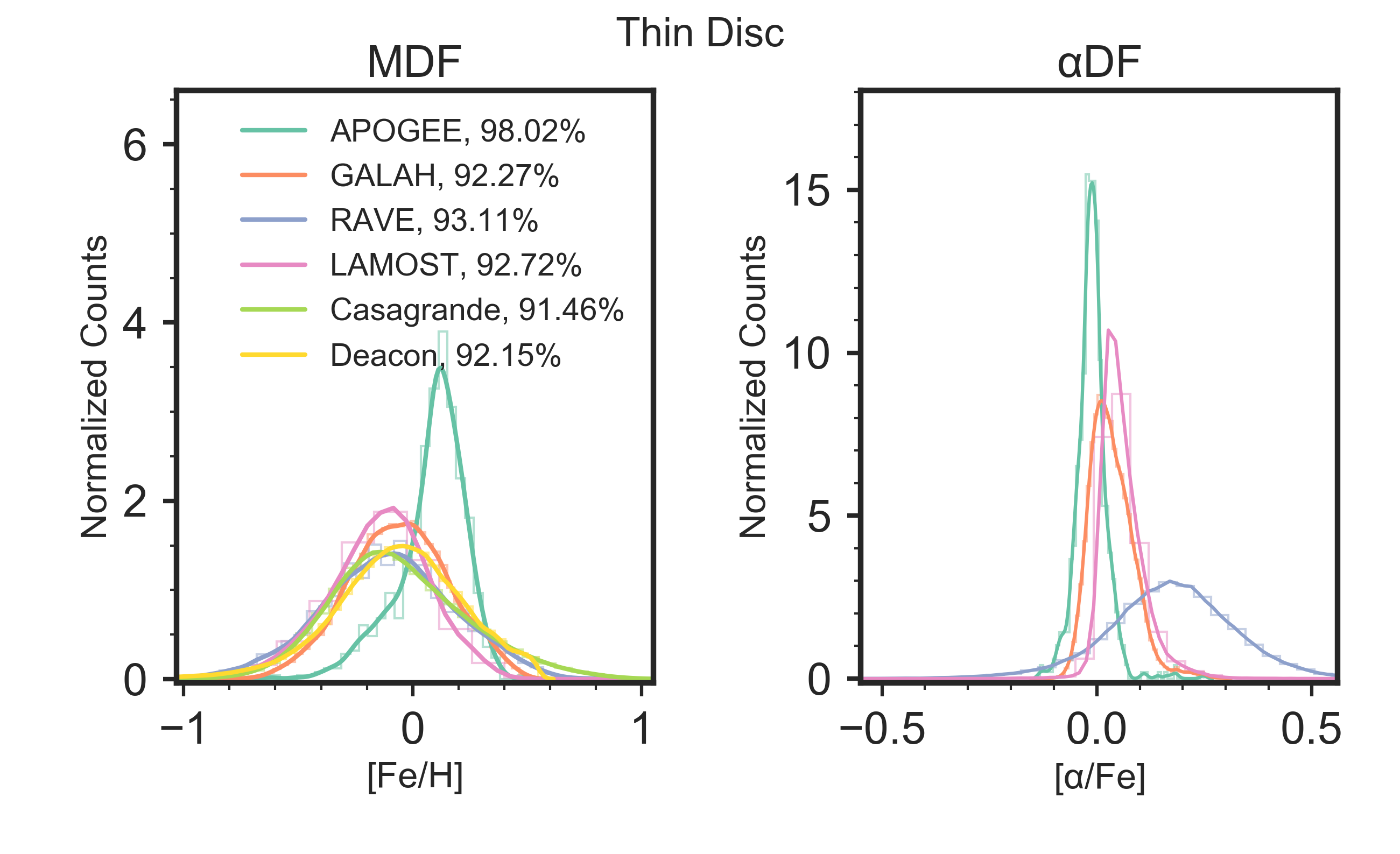}
\includegraphics[width=0.60\textwidth]{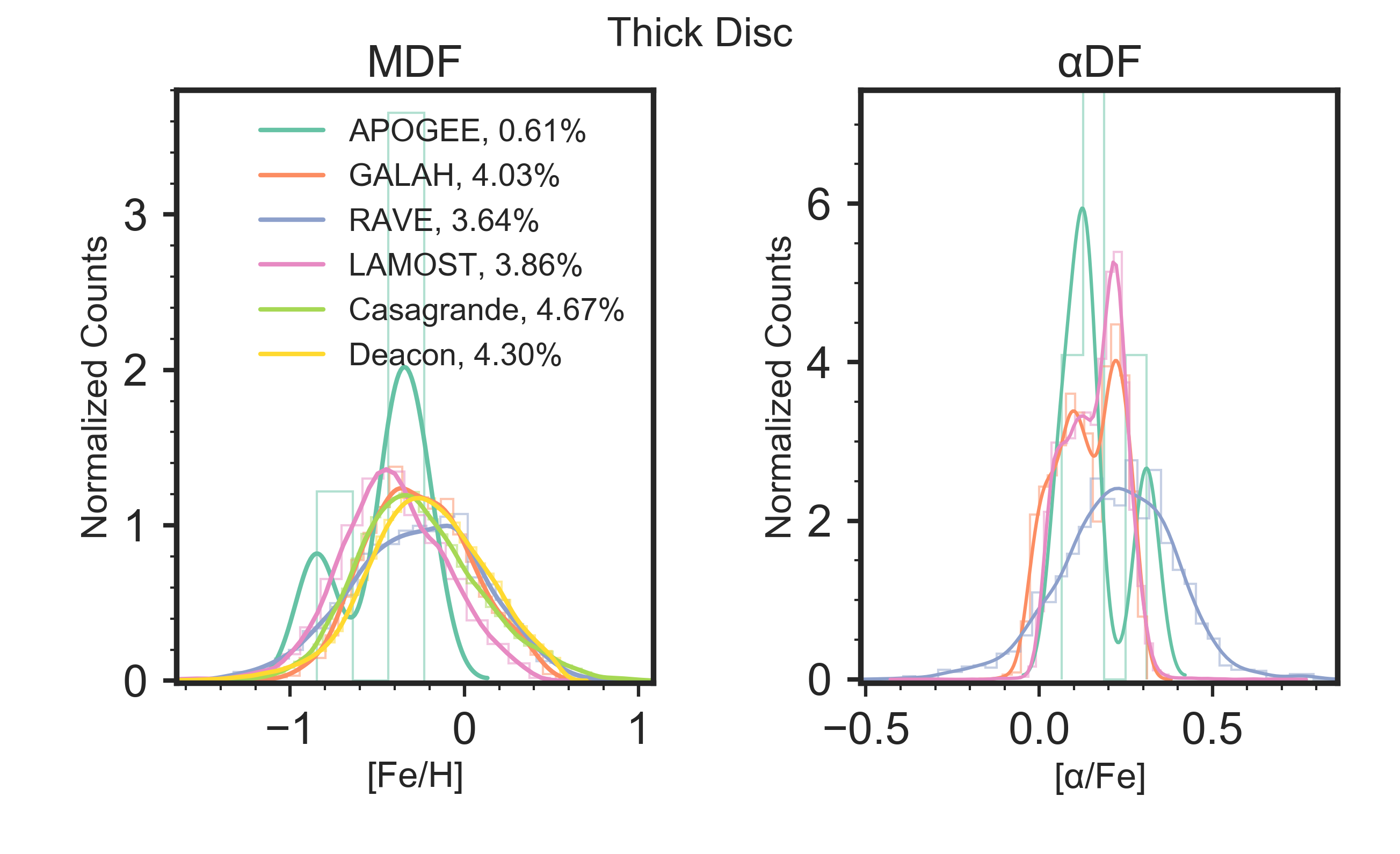}
\includegraphics[width=0.60\textwidth]{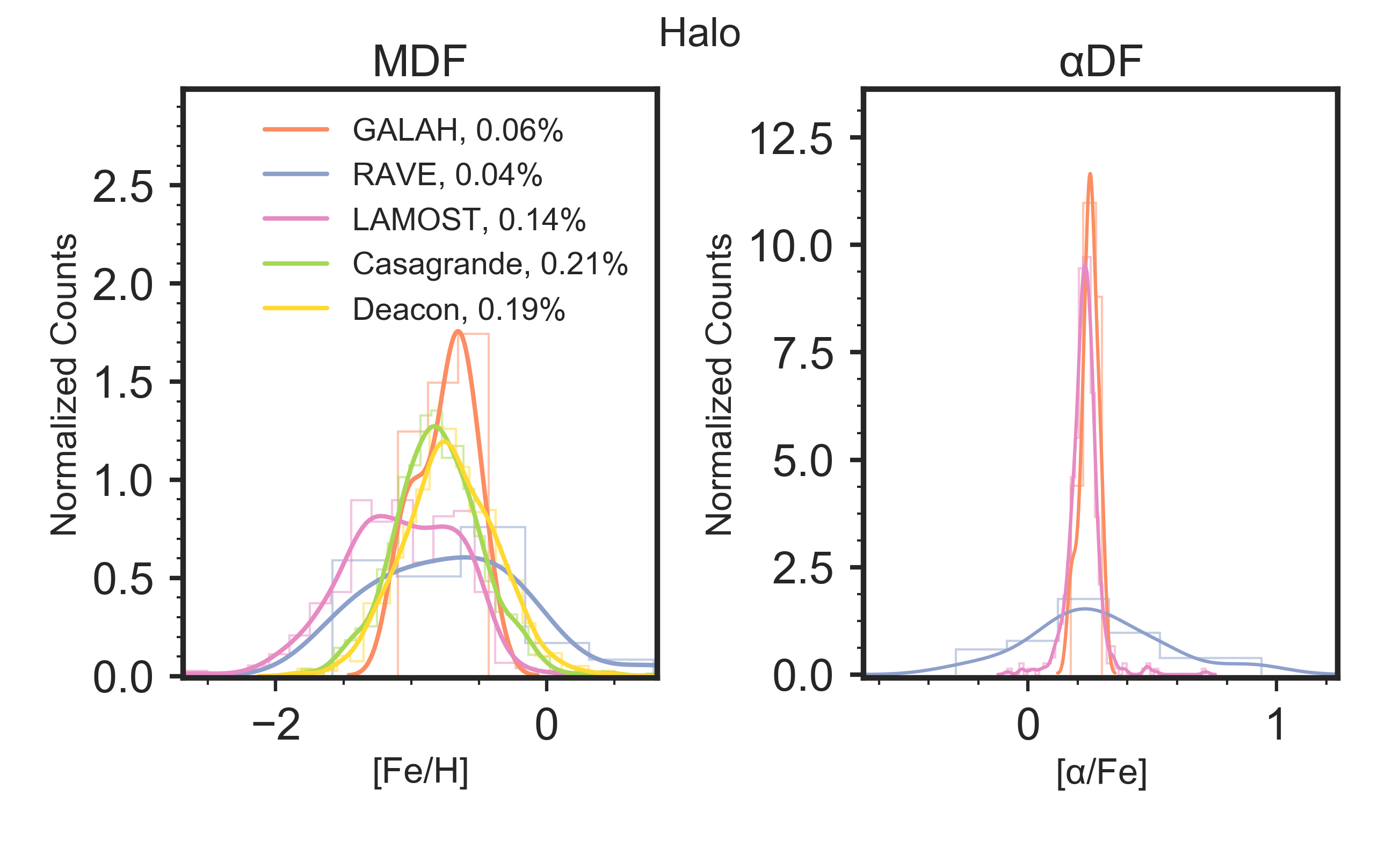}
\caption{MDF and $\alpha$DF for the kinematically-defined thin disc (top left), thick disc (top right), and halo (bottom) as defined in Section \ref{sec:kinematics}. The MDF peaks at lower values while the  $\alpha$DF peaks at higher values for all surveys as one goes from the thin disc and to the halo. }
\label{fig:mdf_adf_components}
\end{figure*}

We explore the kinematics of \textit{TESS} host stars and investigate their chemistry using the spectroscopic and photometric surveys.  In Figure \ref{fig:kin_chem_FeH_spec}, we use the spectroscopic and photometric metallicities to colour-code the cross-matched stars in the Toomre diagrams. Note that the range of the colourbars i.e., [Fe/H] is different for each survey. The Toomre diagrams of each survey show that stars belonging to the thin disc, thick disc, and halo have different characteristic metallicities. We also use the [$\alpha$/Fe] from APOGEE, GALAH, RAVE, and LAMOST to represent a third quantity in the Toomre diagrams shown in Figure \ref{fig:kin_chem_alpha}. Similar to Figure \ref{fig:kin_chem_FeH_spec}, the different Galactic components show varying characteristic  [$\alpha$/Fe] except for RAVE, as discussed in Section \ref{sec:chemistry}. 

We analyze the chemistry of \textit{TESS} host stars for different kinematic memberships. In Figure \ref{fig:mdf}, we demonstrated that in general, the MDFs peak just slightly above or below solar metallicity for all the surveys. The MDFs also show a long tail towards lower metallicities. These results confirm that the majority of TESS CTL stars are nearby and mostly part of the thin disc, but that there are also stars from the thick disc and halo at lower metallicities. Using the same membership criteria from Section \ref{sec:kinematics}, we plot the MDFs and [$\alpha$/Fe] distribution functions ($\alpha$DF) for each kinematically-defined Galactic component -- thin disc, thick disc, and halo -- in Figure \ref{fig:mdf_adf_components} using metallicities and [$\alpha$/Fe] (where available) from APOGEE, GALAH, RAVE, LAMOST, \citet{casagrande19}, and \citet{deacon19}.

\subsubsection{Thin Disc}
The fraction of the TG cross-matched thin disc stars in each survey is $\sim$90\%, which is similar to that of the complete TG except for APOGEE (98\%, see Figure \ref{fig:mdf_adf_components}).  The metallicities for the thin disc stars in Figure \ref{fig:kin_chem_FeH_spec}, shown as diamonds concentrated around (0,0), are higher compared to the rest of the TG cross-matched sample. This is also seen in the thin disc MDF in Figure \ref{fig:mdf_adf_components} where all the surveys peak at around [Fe/H] =  0$\pm 0.2$ dex.  APOGEE shows the lowest dispersion in MDF (though highly negatively skewed), followed by LAMOST and GALAH, and then by RAVE, \citet{casagrande19}, and \citet{deacon19}.

We also show [$\alpha$/Fe] when available (APOGEE, GALAH, RAVE, and LAMOST).  The [$\alpha$/Fe] values for the thin disc stars are the lowest of all the Galactic components as shown in Figure \ref{fig:kin_chem_alpha}.   This is confirmed by the $\alpha$DF in Figure \ref{fig:mdf_adf_components} which shows that both APOGEE and GALAH peak at [$\alpha$/Fe] = 0 dex and are positively skewed. RAVE, on the other hand, is more Gaussian-shaped, peaks at [$\alpha$/Fe]  =  0.20 dex, and extends to both higher and lower [$\alpha$/Fe] values compared to APOGEE and GALAH. LAMOST is somewhere in the middle, peaking at [$\alpha$/Fe] = 0.07 dex. 

We use the [$\alpha$/Fe] vs [Fe/H] diagram to distinguish the estimated contamination fraction of chemical thick disc stars to the kinematic thin disc following the works of \citet{weinberg18} and \citet{buder19} for APOGEE and GALAH, respectively. We also estimated this dividing line by eye for the LAMOST data. We show in Figure \ref{fig:alphaFe} the dividing lines for the thin disc and thick disc that we use for this analysis. Only 1\% of the kinematic thin disc seem to be chemical thick disc stars for APOGEE , while it is 12\% for GALAH and 13\% for LAMOST. We caution that disentangling the two populations is not fool proof as they overlap at higher metallicities. We also did not do this for the TG cross-matched with RAVE because of its uncertainties that blur the distinction between the two populations. 

We now compare the TG kinematic thin disc's chemistry to abundances in the literature. \citet{hayden15} shows the MDF and $\alpha$DF for APOGEE DR12 stars in different Galactic locations that have varying contributions from the chemically-defined thin disc and thick disc stars. Their MDF and $\alpha$DF that best match what we see for our kinematic thin disc stars are those within the solar neighborhood i.e. $7<R<9$ kpc and in the plane of the disc. 
\citet{buder19} used GALAH to make $\alpha$DFs for stars in different metallicity bins (see their Figure 13) and show that for the higher metallicity bins with the greater contribution from thin disc stars (i.e. [Fe/H] $> -0.45$ dex), [$\alpha$/Fe] has a primary peak at $\rm 0 < [\alpha/Fe] < 0.13$ dex and a secondary peak at $\rm 0.13 < [\alpha/Fe] < 0.2$ dex, which we do not see in our GALAH $\alpha$DF but is hinted at by the estimated contamination fraction. 

Other authors have defined the thin disc with alternative ways. For example, \citet{boeche13} used RAVE and assigned component-membership based on a $Z_{max}$ vs $e$ diagram, where $Z_{max}$ is a star's maximum distance from the Galactic plane and $e$ is its orbital eccentricity,  with thin disc stars having low values for both quantities. Their thin disc MDF peaks at a lower metallicity ([Fe/H] = -0.20 dex) compared to stars from APOGEE, GALAH, LAMOST, and \citet{deacon19}. They also show a $\alpha$DF where the thin disc component peaks at $\sim$0.20 dex, similar to what we find for the TG sample cross-matched with RAVE, but not to the TG sample cross-matched with APOGEE, GALAH, and LAMOST. \citet{wang19lamost} added an important quantity in Galactic component separation, age, and used LAMOST to examine the MDF and $\alpha$DF for similarly-aged populations in different Galactic locations. Comparing their thin disc MDFs to ours, their stars younger than 8 Gyr and between $6<R<9$ kpc show the most resemblance to our kinematically-defined thin disc MDF while only the stars at heights $>~0.3$ kpc and younger than 8 Gyr at the same Galactocentric radii show the same trends as our $\alpha$DF. 

In addition to observations, it is also important to contrast our results with simulations. Comparison with models from \citealt{minchev14} (see their Figure 9), who tested whether migration of stars explains why we see the changing skewness in the MDF, show that our thin disc chemistry is consistent with their sample within $7<R<11$ kpc and are confined to the smallest scale height.  

From these comparisons, it is evident that the TG kinematic thin disc sample mainly probes stars in the Solar neighborhood i.e. closer to the plane of the disc and at similar Galactocentric radii as the Sun. Since the thin disc in all TG cross-matched surveys peak at higher metallicity than the thick disc and halo, we would expect to discover more gas giants around stars that belong to this component \citep{fischervalenti05}.

\subsubsection{Thick Disc}
The TG cross-matched thick disc stars from each survey again show similar percentages compared to the whole TG sample (4\%) except for APOGEE which has a lower percentage of thick disc stars at 0.61\%. The thick disc stars (plus) in  Figure \ref{fig:kin_chem_FeH_spec} show a lower range of metallicities compared to thin disc stars (diamond) as is expected.  This is also evident in the thick disc MDF in Figure \ref{fig:mdf_adf_components} where all the surveys peak between -0.9$ <$ [Fe/H] $<$-0.3 dex. 

We also explore the [$\alpha$/Fe] of the thick disc stars in Figure \ref{fig:kin_chem_alpha} that show qualitatively higher [$\alpha$/Fe] values than the thin disc stars. From Figure \ref{fig:kin_chem_alpha}, the $\alpha$DF for APOGEE has two peaks at [$\alpha$/Fe]  = 0.23, and 0.35 dex,  with the lower peak having the higher amplitude. For GALAH, there are two peaks in the $\alpha$DF, one at 0.1 dex and another at 0.25 dex with the peak at 0.25 dex being slightly higher. This is similar to LAMOST's that show two peaks at [$\alpha$/Fe] = 0.1 and 0.25 dex. RAVE's  $\alpha$DF peaks at [$\alpha$/Fe] = 0.25 dex, quite similar to the $\alpha$DF for the thin disc stars in TG cross-matched with RAVE. The multiple peaks in the $\alpha$DFs suggest the presence of multiple stellar populations. Following the contamination determination in Section 4.1.1, the chemical thin disc contamination to the kinematic thick disc is 50\% for APOGEE, 37\% for GALAH, and 36\% for LAMOST. We caution that the kinematic thick disc for the TG crossmatched with APOGEE sample contains only four stars and therefore suffers from small number statistics.In addition, the GALAH and LAMOST thick disc contamination fractions are overestimated because of how we have distinguished a chemical thick disc star from a chemical thin disc star i.e., they becomes less distinct at higher metallicities.

Our thick disc MDFs are consistent with the MDFs of stars at the largest scale height from \citet{hayden15} which largely probe the thick disc. The MDF and $\alpha$DF of simulated thick disc stars from \citet{minchev14} are consistent with kinematic thick disc stars in this study, except for the fact that they only see the higher peak in [$\alpha$/Fe] compared to the multiple peaks that we see in APOGEE, GALAH, and LAMOST. Other studies have sampled stars at higher scale heights that better probe the thick disc and halo regions. \citet{liu18} used LAMOST to study a sample of stars with $|z| > 5$ kpc and constructed MDFs to see the contributions from the thick disc, inner halo, and outer halo. Their thick disc MDFs at all heights are consistent with what we see for our MDFs and peak at [Fe/H] = -0.5 dex.

We have shown that the TG cross-matched kinematic thick disc is lower in metallicity but higher in [$\alpha$/Fe] than the thin disc, similar to comparisons with thick disc in the literature. Planets around low-metallicity, thick disc stars have indeed been found (e.g. \citealt{mayor04,cochran07}). These thick disc stars' elevated [$\alpha$/Fe] abundances may have repercussions for planet-formation, as it is typically the $\alpha$-elements, especially Silicon, that contribute the most to the cores of gas giants \citep{brugamyer11}.

\subsubsection{Halo}

Lastly, we investigate the halo stars that make up $<$1\% of the TG sample.  All of the cross-matched spectroscopic and photometric surveys show similar ($\rm < 1\%$) contribution from halo stars. There is only one cross-matched halo star from APOGEE which we decided not to include in the MDF and $\alpha$DF. The halo stars (triangle) in  Figure \ref{fig:kin_chem_FeH_spec} show the lowest range of metallicities among the Galactic components as well as the largest spread in $\sqrt{U^{2} + W^{2}}$ with no ordered rotation.

This range in metallicity is further supported by the halo MDF in Figure \ref{fig:mdf_adf_components} where all the surveys peak between -1.5$ <$ [Fe/H] $<$-0.8 dex. The MDF peaks for all six surveys also vary from one another. LAMOST shows two metallicity peaks ([Fe/H] = -1.4 and -0.7 dex) while RAVE,  GALAH,  \citet{deacon19}, and \citet{casagrande19} show one peak at around [Fe/H] = -0.8 dex. The multiple peaks for LAMOST may be attributed to halo sub-populations with different origins \citep{carollo10,nissen10,bensby14,belokurov18,helmi18}.

We also examine the [$\alpha$/Fe] of the halo stars using the TG sample cross-matched with APOGEE, GALAH, RAVE, and LAMOST. The Toomre diagram from  Figure \ref{fig:kin_chem_alpha} show that the lone halo star in TG cross-matched with APOGEE has high  [$\alpha$/Fe] compared to the thin disc and thick disc. The TG cross-matched with GALAH and LAMOST halo stars also show the highest [$\alpha$/Fe] compared to the thin disc and thick disc. RAVE on the hand other shows stars with both  high and low [$\alpha$/Fe] abundances. The range in [$\alpha$/Fe] values is shown better in the halo $\alpha$DFs in Figure \ref{fig:kin_chem_alpha}. GALAH, RAVE, and LAMOST all peak at  [$\alpha$/Fe] =  0.25 dex with RAVE showing a second peak at around  [$\alpha$/Fe] =  0.9 dex. We are not able to determine the contamination fraction of chemically thick disc stars to the halo as the switch from the thick disc to the halo in the [$\alpha$/Fe] vs [Fe/H] diagram is quite smooth.

We again compare our results to previous work. From \citet{boeche13}, for stars that have the highest $Z_{max}$ and $e$ --- i.e. have halo kinematics (panel i in their Figures 4 and 11) --- the MDF and $\alpha$DF in their Figure 12 are similar to what we find for the TG sample cross-matched with GALAH, \citet{casagrande19}, and \citet{deacon19}. That is, \citet{boeche13} have halo stars peaking at [Fe/H] $\approx$ -1.0 dex and at [$\alpha$/Fe] $\approx$ 0.25 dex. \citet{an15} used $ugriz$ photometry from SDSS Strip 82 to investigate the Milky Way halo. They constructed MDFs that show two peaks at [Fe/H] = -1.4 dex and -1.9 dex that they attribute to an inner halo and an outer halo with prograde and retrograde motions, respectively. This is similar to what we see for LAMOST, though shifted to higher metallicities. \citet{liu18} also find an inner halo and outer halo MDF peak at [Fe/H] = -1.2 dex and -1.9 dex, respectively, using LAMOST. 

\subsection{The Chemo-kinematics of the \textit{TESS} Objects of Interest}
\label{sec:discoveries}

\begin{figure}
\center
\includegraphics[width=0.4\textwidth]{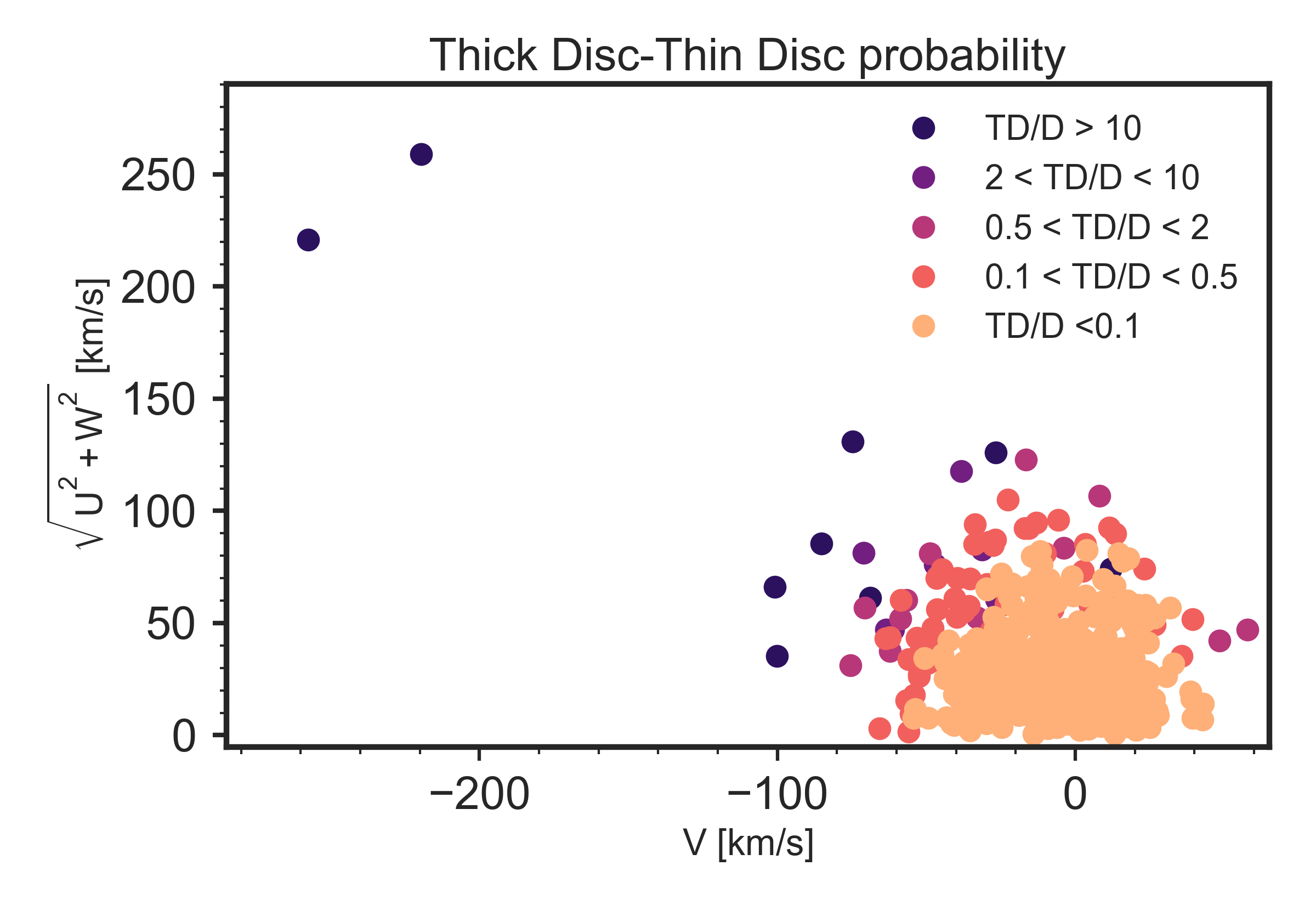}
\includegraphics[width=0.4\textwidth]{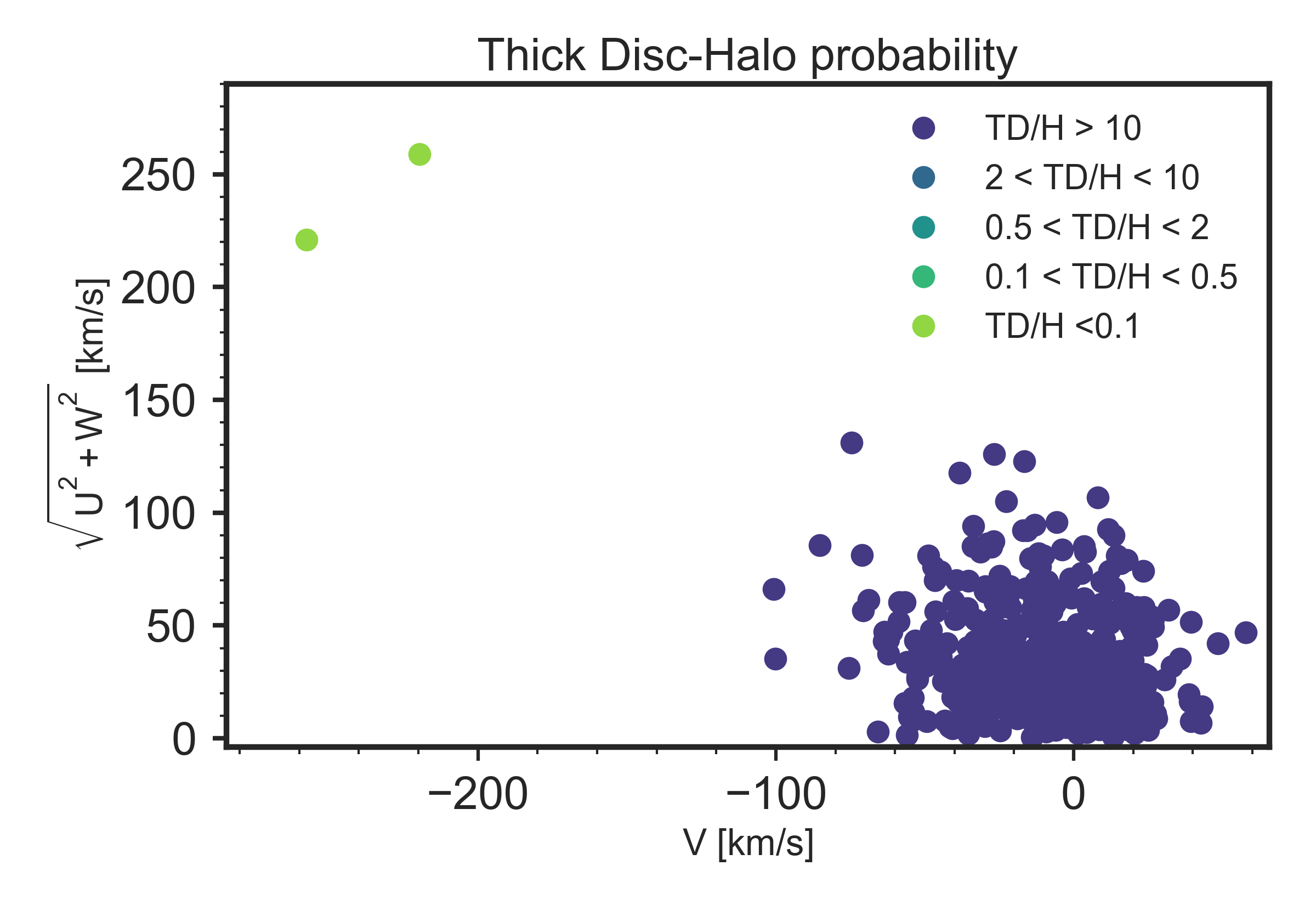}
\caption{Top panel: Toomre diagram for the 563 \textit{TESS} Objects of Interest (TOI) in common with TGv8 colour-coded by the thick disc to thin disc probability, where purple ($TD/D > 10$) denotes that a star is 10 times more likely to be part of the thick disc than the thin disc. Bottom panel: Toomre diagram for the same sources colour-coded by the thick disc to halo membership where green ($TD/H < 0.1$) represents a star that is 10 times more likely to be part of the halo than the thick disc.}
\label{fig:toomre_toi}
\end{figure}

The \textit{TESS} Objects of Interest (TOI) are sources that \textit{TESS} has found to be possible planet-hosting stars, and are therefore good targets for follow-up observations. We obtained the TOI from \url{https://exofop.ipac.caltech.edu/tess/view_toi.php}\footnote{8 October 2019 version} that has 1,183 planet candidates. We cross-matched their host stars with the sources in TGv8 to determine their chemo-kinematics and end up with 563 sources. Figure \ref{fig:toomre_toi} shows the Toomre diagrams colour-coded by thick disc-thin disc probability and thick disc-halo probability for the 563 TOI sources, each clearly demonstrating that most of the TOIs are in the thin disc. Using the same component membership criteria outlined in Section \ref{sec:kinematics}, we find that 94.85\% (534 stars) of the TOIs belong to the thin disc, 2.66\% (15 stars) belong to the thick disc, and 0.36\% (2 stars, TOI 152 and TOI 320) belong to the halo. These stellar population fractions are similar to what we find for the whole TGv8 sample with the majority of the sources being in the thin disc.


As of 08 October 2019, there were 29 confirmed exoplanets discovered with \textit{TESS}. 
 As expected, most of these confirmed exoplanet-hosting stars are in the thin disc (e.g. \citealt{jones18,wang18,dragomir19,huber19}). \citet{esposito18} reported the existence of a hot Neptune orbiting an inactive G7 dwarf star, TOI 118, which we classify as a kinematic thick disc star. TOI 118 is also cross-matched with  \citet{casagrande19} and \citet{deacon19}, where the reported metallicities are $0.341 \pm 0.171$ dex and $0.359^{+0.177}_{-0.185}$ dex, respectively, different from the spectroscopic metallicity from \citet{esposito18} i.e. 0.04 $\pm$ 0.04 dex. None of the confirmed exoplanet-hosting stars reside in the halo, to date.

We expect this catalogue to be more applicable for statistical studies as more exoplanets candidates discovered with \textit{TESS} are confirmed. Recent work on the Galactic context of exoplanet populations has been made available utilizing \textit{Kepler} data. For example, \citet{mctier19} examined 1647 stars with confirmed planets from \textit{Kepler} and measured lower velocities for these host stars than the average for the field sample, which may be an evidence of the planet-metallicity correlation or thick/thin disc membership. However, comparisons to non-\textit{Kepler} host stars with identical properties reveal that planets are just as likely to form around fast stars as they do around slow stars, and that the observed velocity difference is an effect of \textit{Kepler}'s selection function in that \textit{Kepler} selects younger, brighter, hotter, and therefore slower stars. This would be interesting to further explore using the \textit{TESS} catalogue outlined in this work, as \textit{TESS} has a different selection function and is expected to discover more exoplanets. 

\section{Summary}
\label{sec:summary}
\textit{TESS} is going to revolutionize exoplanet studies by discovering thousands of planets around the nearest and brightest stars. Yet the chemo-kinematic properties of these stars have yet to be explored. We have constructed a catalogue of  2,218,434 \textit{TESS}  CTL targets with chemo-kinematic properties from $Gaia$ DR2, APOGEE, GALAH, RAVE, LAMOST, and photometrically-derived stellar parameter catalogues from \citet{casagrande19} and \citet{deacon19}.  We compute thin disc, thick disc, and halo membership probabilities for the stars in our catalogue based on their kinematics inferred from $Gaia$, following \citet{bensby14}. While the  majority of \textit{TESS} CTL targets are in the thin disc, we find that 4\% of them ($\sim$89,000 stars) reside in the thick disc and $<1\%$ of them ($\sim$ 22,000 stars) are in the halo. We find similar percentage for the different Galactic components using other cross-matched surveys with the exception of APOGEE which primarily targets more distant red giants that are preferentially removed for TIC/CTL v8. The TOIs in common with the TG sample are also comprised of similar fractions of the thin disc, thick disc, and halo with a majority of them being in the thin disc.

We explore the kinematics of the thin disc, thick disc, and halo stars and show that for all cross-matched surveys, there is decreasing metallicity and increasing [$\alpha$/Fe] from the thin disc to the halo as seen in the Toomre diagrams. We also explore the metallicity and [$\alpha$/Fe] distributions of each Galactic component and confirm that the  MDF peak moves to lower values and the $\alpha$DF peak moves to higher values as one goes from the thin disc to the halo. 

The catalogue we have generated will be advantageous to use for and valuable in curating samples based on kinematics, kinematic component-membership, metallicity, or [$\alpha$/Fe]. In the era of big data, especially relating to the search for exoplanets, it is now possible and in fact imperative to look at all these dimensions and properties; on how kinematics of host stars relate to their component-membership, how these component memberships relate to the host star metallicity and chemical abundance, how the metallicity and chemical abundance relate to the available materials for planet formation, and ultimately, how all of these tie together to produce the trends in the planet populations that we observe and that we expect to see.

\section*{Acknowledgements}
\begin{footnotesize}
This work was performed in part at Aspen Center for Physics, which is supported by National Science Foundation grant PHY-1607611. AC thanks the LSSTC Data Science Fellowship Program, which is funded by LSSTC, NSF Cybertraining Grant 1829740, the Brinson Foundation, and the Moore Foundation. KH is partially supported by a Research Corporation TDA/Scialog grant. KH acknowledges support from the National Science Foundation grant AST-1907417. BPB acknowledges support from the National Science Foundation grant AST-1909209. 
This paper includes data collected by the TESS mission, which are publicly available from the Mikulski Archive for Space Telescopes (MAST). Funding for the TESS mission is provided by NASA's Science Mission directorate. This work has made use of data from the European Space Agency (ESA)
mission {\it $Gaia$} (\url{https://www.cosmos.esa.int/gaia}), processed by
the {\it $Gaia$} Data Processing and Analysis Consortium (DPAC,
\url{https://www.cosmos.esa.int/web/gaia/dpac/consortium}). Funding
for the DPAC has been provided by national institutions, in particular
the institutions participating in the {\it $Gaia$} Multilateral Agreement. Funding for the Sloan Digital Sky Survey IV has been provided
by the Alfred P. Sloan Foundation, the U.S. Department of Energy Office of Science,
and the Participating Institutions. SDSS acknowledges support and resources from the
Center for High-Performance Computing at the University of Utah. The SDSS web site
is \url{www.sdss.org}.
SDSS is managed by the Astrophysical Research Consortium for the Participating
Institutions of the SDSS Collaboration including the Brazilian Participation Group, the Carnegie Institution for Science, Carnegie Mellon University, the Chilean Participation Group, the French Participation Group, Harvard-Smithsonian Center for Astrophysics, Instituto de Astrofasica de Canarias, The Johns Hopkins University, Kavli Institute for the Physics and Mathematics of the Universe (IPMU) / University of Tokyo,
Lawrence Berkeley National Laboratory, Leibniz Institut f\"ur Astrophysik Potsdam
(AIP), Max-Planck-Institut f\"ur Astronomie (MPIA Heidelberg), Max-Planck-Institut f\"ur
Astrophysik (MPA Garching), Max-Planck-Institut f\"ur Extraterrestrische Physik (MPE),
National Astronomical Observatory of China, New Mexico State University, New York
University, University of Notre Dame, Observat\'orio Nacional / MCTI, The Ohio State University, Pennsylvania State University, Shanghai Astronomical Observatory, United
Kingdom Participation Group, Universidad Nacional Aut\'onoma de M\'exico, University
of Arizona, University of colourado Boulder, University of Oxford, University of
Portsmouth, University of Utah, University of Virginia, University of Washington, University of Wisconsin, Vanderbilt University, and Yale University. This work has made use of GALAH DR2, based on data acquired through the Australian Astronomical Observatory, under programmes: A/2013B/13 (The GALAH pilot survey);
A/2014A/25, A/2015A/19, A2017A/18 (The GALAH survey). We acknowledge the traditional owners of the land on which the AAT stands, the Gamilaraay people, and pay our respects to elders past and present. Funding for RAVE has been provided by: the Australian Astronomical Observatory; the Leibniz-Institut fuer Astrophysik Potsdam (AIP); the Australian National University; the Australian Research Council; the French National Research Agency; the German Research Foundation (SPP 1177 and SFB 881); the European Research Council (ERC-StG 240271 Galactica); the Istituto Nazionale di Astrofisica at Padova; The Johns Hopkins University; the National Science Foundation of the USA (AST-0908326); the W. M. Keck foundation; the Macquarie University; the Netherlands Research School for Astronomy; the Natural Sciences and Engineering Research Council of Canada; the Slovenian Research Agency; the Swiss National Science Foundation; the Science \& Technology Facilities Council of the UK; Opticon; Strasbourg Observatory; and the Universities of Groningen, Heidelberg and Sydney. The RAVE web site is at \url{https://www.rave-survey.org}. Guoshoujing Telescope (the Large Sky Area Multi-Object Fiber Spectroscopic Telescope LAMOST) is a National Major Scientific Project built by the Chinese Academy of Sciences. Funding for the project has been provided by the National Development and Reform Commission. LAMOST is operated and managed by the National Astronomical Observatories, Chinese Academy of Sciences.
\end{footnotesize}

\bibliographystyle{mnras}
\bibliography{references} 




\appendix
\section{\textit{TESS} Input Catalogue v7}
\label{sec:TIC7}

Although this study was done using TIC v8, we also explore TIC v7 \citep{stassun18}. The main difference between v8 and v7 is that v8 uses \textit{Gaia} DR2 as base, compared to TIC v7 that uses 2MASS. Going from the v7 to v8 changes the estimated \textit{T} mag, \teff, stellar mass, and stellar radius, as well as increasing the number of sources with these estimated properties. For more details on TIC/CTL v7, we refer the reader to \citet{stassun18}.

We did the same analysis to TIC/CTL v7 as in Section \ref{sec:data} and call the TESS cross-matched with \textit{Gaia} table TGv7. Below we list the main differences between TGv8 and TGv7: 
\begin{itemize}
    \item {TGv7 has 1,405,352 sources while TGv8 has 2,218,434 sources.}
    \item {The CMD of TGv7 has more contamination from subgiant and red giant branch stars which was improved upon by TGv8 because of the addition of \textit{Gaia} DR2 information.}
    \item{The breakdown for kinematic component-membership is slightly different and has changed as follows: thin disc: from 91.77\% in TGv7 to 93.17\% in TGv8; thick disc: from 4.42\% to 3.64\%; halo: from 0.50\% to 0.16\%; thin-thick disc transition: from 3.25\% to 3.00\%; thick disc-halo transition: from 0.06\% to 0.03\%.}
    \item {The cross-matched TGv7 with other surveys yielded different number of sources compared to TGv8 (APOGEE: from 2,174 to 658; GALAH: from 19,030 to 32,517; RAVE: from 55,509 to 59,984; LAMOST: from 237,739 to 344,565; \citet{casagrande19}: from 356,936 to 500,007; \citet{deacon19}: from 424,504 to 413,100).}
    \item{There is a larger fraction of thick disc and halo stars in TGv7 compared to TGv8, though these stars may be too dim for follow-up observations.}
\end{itemize}

\onecolumn
\begin{center}
\section{Online Table}
\begin{longtable}{p{1cm} p{2.5cm} p{1.25cm} p{1.25cm} p{8.75cm}} 
\caption{Catalogue of \textit{TESS} stellar properties from this work}\\
\hline 
 Column & Label & Format & Units & Notes\\
\hline
  1 & \textit{Gaia} source\_id & Long & & \\
  2 & \textit{Gaia} RA & Double & deg & \\  
  3 & \textit{Gaia} Dec & Double & deg & \\ 
  4 & \textit{Gaia} parallax & Double & mas & \\
  5 & \textit{Gaia} e\_parallax & Double & mas & \\
  6 & \textit{Gaia} pmra & Double & mas $\rm yr^{-1}$ & \\
  7 & \textit{Gaia} e\_pmra & Double & mas $\rm yr^{-1}$ & \\
  8 & \textit{Gaia} pmdec & Double & mas $\rm yr^{-1}$ & \\
  9 & \textit{Gaia} e\_pmdec & Double & mas $\rm yr^{-1}$ & \\
  10 & \textit{Gaia} vrad & Double & km $\rm s^{-1}$ & \\  
  11 & \textit{Gaia} e\_vrad & Double & km $\rm s^{-1}$ & \\ 
  12 & \textit{Gaia} G mag & Double & mag & \textit{Gaia} \textit{G} band apparent magnitude\\  
  13 & \textit{Gaia} G mag error & Double & mag & \\  
  14 & \textit{Gaia} BP mag & Double & mag & \textit{Gaia} blue pass band apparent magnitude\\  
  15 & \textit{Gaia} BP mag error & Double & mag & \\  
  16 & \textit{Gaia} RP mag & Double & mag & \textit{Gaia} red pass band apparent magnitude\\  
  17 & \textit{Gaia} RP mag error & Double & mag & \\ 
\hline  
  18 & Marchetti U & Double & km $\rm s^{-1}$ & Cartesian Galactocentric $x$ velocity \\   
  19 & Marchetti el\_U & Double & km $\rm s^{-1}$ & Lower uncertainty on Cartesian Galactocentric $x$ velocity \\ 
  20 & Marchetti eu\_U & Double & km $\rm s^{-1}$ & Upper uncertainty on Cartesian Galactocentric $x$ velocity \\ 
  21 & Marchetti V & Double & km $\rm s^{-1}$ & Cartesian Galactocentric $y$ velocity \\   
  22 & Marchetti el\_V & Double & km $\rm s^{-1}$ & Lower uncertainty on Cartesian Galactocentric $y$ velocity \\ 
  23 & Marchetti eu\_V & Double & km $\rm s^{-1}$ & Upper uncertainty on Cartesian Galactocentric $y$ velocity \\   
  24 & Marchetti W & Double & km $\rm s^{-1}$ & Cartesian Galactocentric $z$ velocity \\   
  25 & Marchetti el\_W & Double & km $\rm s^{-1}$ & Lower uncertainty on Cartesian Galactocentric $z$ velocity \\ 
  26 & Marchetti eu\_W & Double & km $\rm s^{-1}$ & Upper uncertainty on Cartesian Galactocentric $z$ velocity \\   
  27 & Marchetti vtot & Double & km $\rm s^{-1}$ & Total velocity in Galactic rest frame \\   
  28 & Marchetti el\_vtot & Double & km $\rm s^{-1}$ & Lower uncertainty on Total velocity in Galactic rest frame \\ 
  29 & Marchetti eu\_vtot & Double & km $\rm s^{-1}$ & Upper uncertainty on Total velocity in Galactic rest frame \\    
  30 & Bailer-Jones r\_est & Double & pc & Estimated distance from \citet{bailerjones18} \\   
  31 & Bailer-Jones r\_lo & Double & pc & Lower bound on estimated distance from \citet{bailerjones18} \\  
  32 & Bailer-Jones r\_hi & Double & pc & Upper bound on estimated distance from \citet{bailerjones18} \\  
  \hline
  33 & 2MASS ID & String & & \\
  34 & TICID & Integer & & \\  
  35 &  TESS RA & Double & deg & \\
  36 &  TESS DEC & Double & deg & \\ 
  37 &  GLONG & Double & deg & \\
  38 &  GLAT & Double & deg & \\   
  39 & TESSMAG & Float & mag & \\
  40 & PRIORITY & Double & & Target priority, 1 is highest \\
  41 & HIP & Float &  & $Hipparcos$ ID \\
  42 & TYCHO2 & Float &  & $Tycho-2$ ID \\  
  \hline
  43 & TD\_D & Double & & thick disc to thin disc membership probability based on \citet{bensby14} \\
  44 & TD\_D\_le & Double & & 16th percentile on the thick disc to thin disc membership probability based on \citet{bensby14} \\
  45 & TD\_D\_ue & Double & & 84th percentile on the thick disc to thin disc membership probability based on \citet{bensby14} \\
  46 & TD\_H & Double & & thick disc to halo membership probability based on \citet{bensby14} \\
  47 & TD\_H\_le & Double & & 16th percentile on the thick disc to halo membership probability based on \citet{bensby14} \\
  48 & TD\_H\_ue & Double & & 84th percentile on the thick disc to halo membership probability based on \citet{bensby14} \\
  49 & D & Double & & $(1 + TD/D + ((TD/D)/(TD/H)))^{-1}$  \\
  50 & TD & Double & & $(1 + TD/H + ((TD/H)/(TD/D)))^{-1}$ \\
  51 & H & Double & & $1 - D - TD$ \\
  \hline
  52 & APOGEE Fe\_H & Double & dex & APOGEE cross-matched with TG sample (see Section \ref{sec:apogee}) \\
  53 & APOGEE Teff & Float & K & \\
  54 & APOGEE logg & Float & dex & \\
  55 & APOGEE alpha & Float & dex & \\
  56 & GALAH Fe\_H & Double & dex & GALAH cross-matched with TG sample (see Section \ref{sec:galah}) \\
  57 & GALAH Teff & Double & K & \\
  58 & GALAH logg & Double & dex & \\ 
  59 & GALAH alpha & Double & dex & \\ 
  60 & RAVE Fe\_H & Double & dex & RAVE cross-matched with TG sample (see Section \ref{sec:rave}) \\
  61 & RAVE Teff & Double & K & \\
  62 & RAVE logg & Double & dex & \\ 
  63 & RAVE alpha & Double & dex & \\ 
  64 & LAMOST Fe\_H & Float & dex & LAMOST cross-matched with TG sample (see Section \ref{sec:lamost}) \\
  65 & LAMOST Teff & Float & K & \\
  66 & LAMOST logg & Float & dex & \\   
  67 & LAMOST alpha & Float & dex & \\   
  68 & Casagrande Fe\_H & Double & dex & \citet{casagrande19} cross-matched with TG sample (see Section \ref{sec:casagrande}) \\
  69 & Casagrande Teff & Short & K & \\
  70 & Deacon Fe\_H & Float & dex & \citet{deacon19} cross-matched with TG sample (see Section \ref{sec:deacon}) \\
  71 & Deacon Teff & Double & K & \\
  72 & Deacon logg & Double & dex & \\
  \hline
  \hline
\label{tab:catalogue}

\end{longtable}

\end{center}



\bsp	
\label{lastpage}
\end{document}